\documentclass[12pt, draftclsnofoot, onecolumn]{IEEEtran}

\usepackage{balance, geometry}
\usepackage{array, calc, color, multicol, verbatim}
\usepackage{algorithm,algorithmicx,algcompatible}
\usepackage{graphics, epstopdf, graphicx, epsfig, psfrag, epsf, subfigure}
\usepackage{amssymb, amsfonts, amsmath, times}
\usepackage{multicol,balance, verbatim}
\usepackage{textcomp, mathrsfs, stmaryrd, setspace}
\usepackage{xspace}
\usepackage{enumerate, url}

\newcommand{\ie}{{\em i.e., }}
\newcommand{\eg}{{\em e.g., }}

\newcommand{\MC}{MicroCast\xspace}
\newcommand{\Requester}{MicroDownload\xspace}
\newcommand{\Distributor}{MicroNC-P2\xspace}
\newcommand{\Broadcast}{MicroBroadcast\xspace}

\newcommand{\Pull}{BitTorrent-Pull\xspace}
\newcommand{\Push}{R2-Push\xspace}

\DeclareGraphicsExtensions{.eps, .pdf, .png, .jpg}   

\newcommand{\Jset}{\mathcal{J}}

\newcommand{\Nset}{\mathcal{N}}
\newcommand{\Hset}{\mathcal{H}}

\begin{document}

\pagenumbering{gobble}

\title{MicroCast: Cooperative Video Streaming \\ using Cellular and D2D Connections}

\author{
     Anh~Le,~\IEEEmembership{Member,~IEEE,}
     Lorenzo~Keller, 
	Hulya~Seferoglu,~\IEEEmembership{Member,~IEEE,}
	Blerim~Cici, 
	Christina~Fragouli,~\IEEEmembership{Member,~IEEE,}
       and~Athina~Markopoulou,~\IEEEmembership{Senior Member,~IEEE}
\thanks{A. Le, B. Cici, and A. Markopoulou are with University of California, Irvine. Emails: {\tt \{anh.le, bcici, athina\}@uci.edu}.  L. Keller is with EPFL. Email: {\tt lorenzo.keller@epfl.ch.} H. Seferoglu is with MIT. Email: {\tt hseferog@mit.edu}. C. Fragouli is with EPFL and UCLA. Email: {\tt christina.fragouli@epfl.ch}}}

\maketitle

\begin{abstract}

We consider a group of mobile users, within proximity of each other, who are interested in watching the same online video at roughly the same time. The common practice today is that each user downloads the video independently on her mobile device using her own cellular connection, which wastes access bandwidth and may also lead to poor video quality. We propose a novel cooperative system where each mobile device uses simultaneously two network interfaces: (i) the cellular to connect to the video server and download parts of the video and (ii) WiFi to connect locally to all other devices in the group and exchange those parts. Devices cooperate to efficiently utilize all network resources and are able to adapt to varying wireless network conditions. In the local WiFi network, we exploit overhearing, and we further combine it with network coding. The end result is savings in cellular bandwidth and improved user experience (faster download) by a factor on the order up to the group size.

We follow a complete approach, from theory to practice. First, we formulate the problem using a network utility maximization (NUM) framework, decompose the problem, and provide a distributed solution. Then, based on the structure of the NUM solution, we design a modular system called {\em MicroCast} and we implement it as an Android application. We provide both simulation results of the NUM solution and experimental evaluation of MicroCast on a testbed consisting of Android phones. We demonstrate that the proposed approach brings significant performance benefits without battery penalty.

\end{abstract} 


\section{\label{sec:intro}Introduction}

Mobile video is already a big part of today's cellular traffic and is expected to grow much faster than other traffic. Indeed, cellular traffic is growing exponentially (tripling every year), with the share of video traffic increasing from 50\% now to an expected 66\% by 2015 \cite{cisco_index}. Credit Suisse reported that in 2012, 23\% of base stations globally have utilization rates of more than 80--85\% in busy hours, up from 20\% in 2011  \cite{credit_suisse}. This dramatic increase in demand poses a challenge on the cellular networks, which are already struggling to provide good services to their subscribers. For example, the data rate of a cellular connection may fluctuate over time (\eg throughout the day); the service loss rate can be as high as 50\% \cite{pc_worldstat}; and coverage can be spotty  depending on the location and user mobility.

Within this important problem space of mobile video, we focus on a particular setting, which we refer to as the {\em ``micro setting''}: we consider a group of mobile users, within proximity of each other, who are interested in watching the same online video via their cellular links  at roughly the same time, as depicted in Fig 1(a). This is a common scenario in practice and examples include the following.  A group of friends  may want to share and watch together a video clip  from YouTube, Netflix, or DailyMotion. In fact, 50\% of YouTube male viewer, who are between 18 and 34 years of age, watch YouTube clips together with friends in person \cite{youtubeV}.
Friends may  want to watch a live soccer match together on their mobile devices while at a remote location, such as a camping or skiing site, where some of the mobile devices may have poor connection. Family members may want to watch the same movie at the same time but each using their own mobile device, \eg while on a train or in a car.  A group of students may want to watch the video of a lecture from an online education system while sitting together and using several mobile devices\footnote{We note that a special case of this scenario is when the video is stored locally on one of the devices (as opposed to online on the server) and the user wants to share it with the other members. Our analysis and implementation are generic enough to address this specific and popular scenario.}.

\color{black}
The default practice today for streaming the same video to a group of users, within proximity of each other, is that each mobile device downloads the video independently from the server using their own cellular connection. This approach has several issues. From a user's perspective, the video download rate  is limited by the individual cellular link rates, which may not be sufficient or stable enough to provide high video quality.  From a cellular provider's perspective, downloading the same content multiple times in the same cell is a waste of precious access bandwidth. Furthermore, many nearby users downloading the same content may lead to cell congestion, which eventually translates to poor user experience as well. What would be desirable from both a user's and a provider's perspectives would be the ability to use the aggregate cellular bandwidth in an efficient way so as to satisfy all users.  The ``micro'' setting has some inherent characteristics  that make it naturally amenable to cooperation, \eg users are engaging in a group activity and are within proximity of each other, thus can establish device-to-device links.

In this paper, we   leverage device-to-device connections between all mobile devices and we cooperatively download the video so as to effectively``bundle up'' the cellular links in the group. More specifically, each mobile device uses simultaneously two network interfaces: (i) the cellular to connect to the video server and download parts of the video and (ii) the WiFi to connect to the rest of the group and exchange those parts. In the local WiFi network, we exploit overhearing and further combine it with network coding to effectively use WiFi resources. Devices cooperate to efficiently utilize all network resources and are able to adapt to varying wireless network conditions. As a result,  the common video download rate can be up to the aggregate of the cellular links of all mobile devices in the group. This benefits both the end-user, who enjoys higher rate and faster download by a factor up to the group size, and  the cellular provider who avoids redundant downloads and saves its access bandwidth by the same factor.

\begin{figure*}[t!]
\centering
  \subfigure[]{\includegraphics[height=50mm]{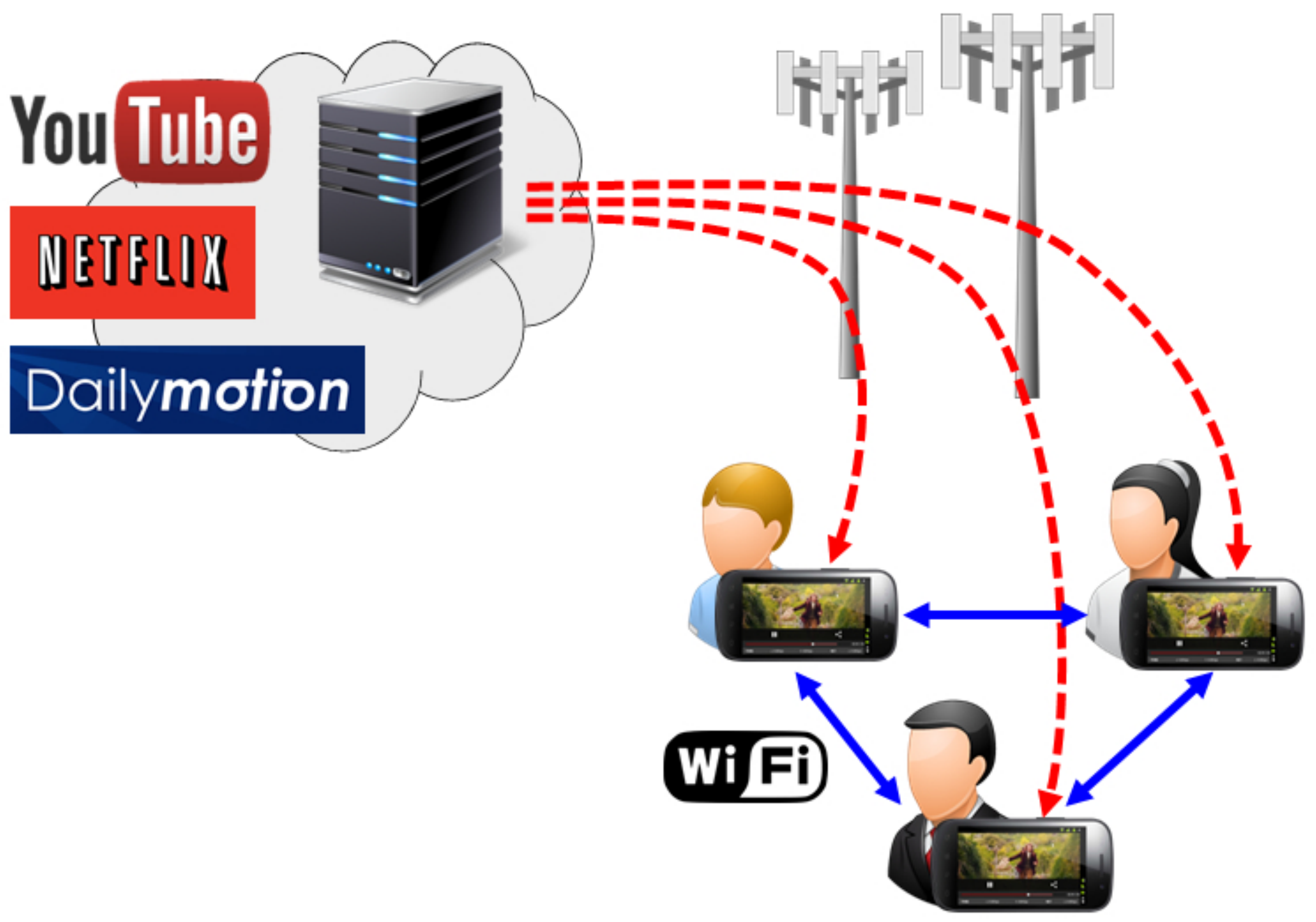}}\hspace*{30mm}
  \subfigure[]{\includegraphics[height=50mm]{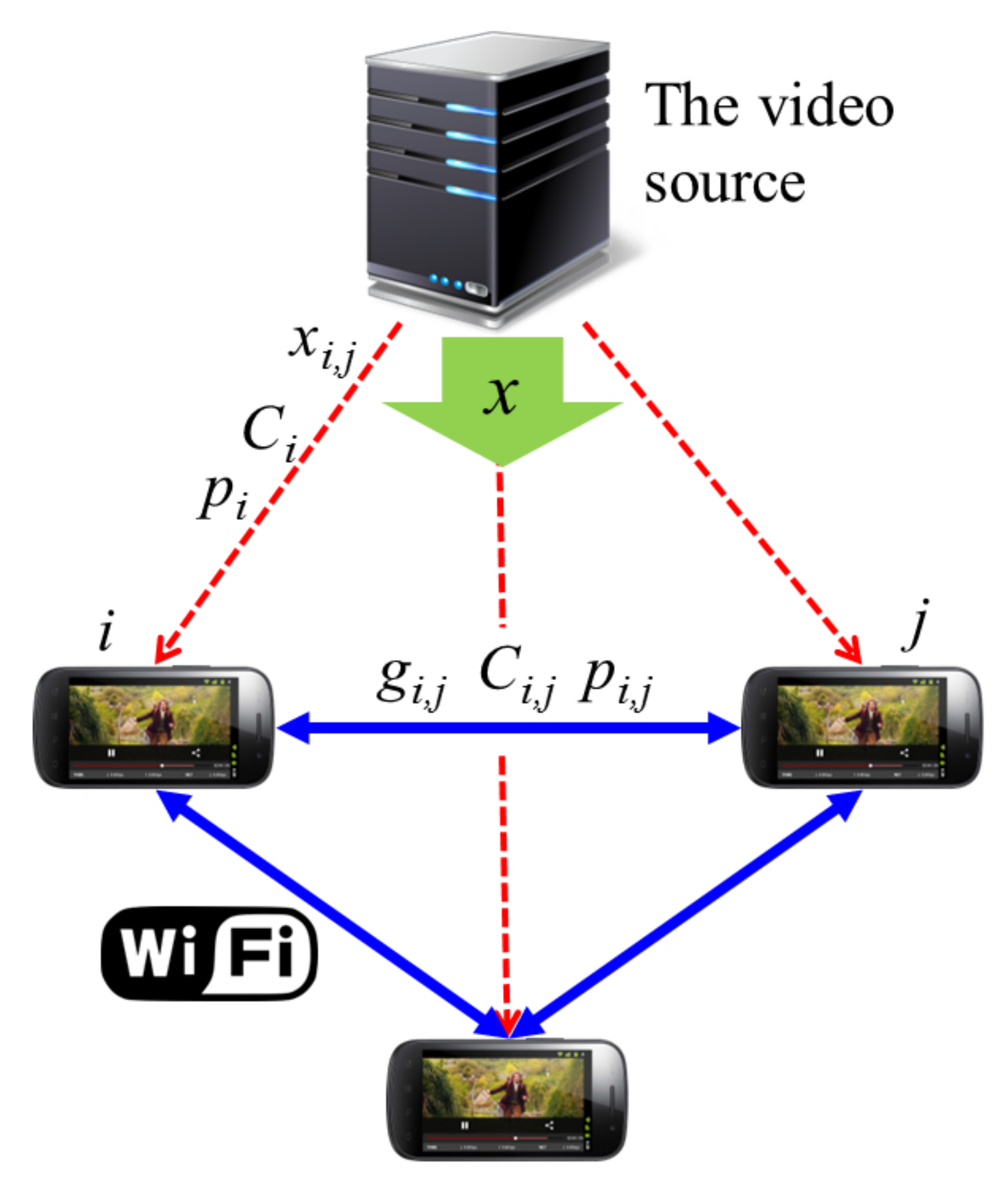}}
\caption{(a) {\em ``Micro'' Setting and MicroCast.} A group of mobile device users, within proximity of each other, is interested in watching the same video at roughly the same time. Each mobile device connects to a video source, \eg YouTube, Netflix, or DailyMotion, using its cellular (3G or 4G) connection. The base stations may be the same or different for different users, depending on the provider they use. Each mobile device can receive packets from the source as well as from other devices in the neighborhood through device-to-device WiFi links. (b) System model used in the analysis (Section IV): the $x$'s are the rates; the $C$'s are the capacities; and the $p$'s are the loss probabilities.
\label{fig:MC-setting-model}}
\vspace{-20pt}
\end{figure*}

More specifically, we take the following steps.  First, we formulate the problem using a network utility maximization (NUM) framework, decompose the problem, and provide a distributed solution. Then, based on the structure of the NUM solution, we design a system called \MC\footnote{{\em Micro} indicates locality: there is a small number of users and they are all within proximity of each other. {\em Cast} indicates a multicast traffic scenario: all users in the group are interested in the same content sent by a single source, and that we utilize WiFi broadcast.}, and we implement a prototype on the Android platform. Fig.~\ref{fig:MC-setting-model} (a) illustrates the micro setting, and Fig. 2(a) depicts the \MC architecture. \MC consists of three main components: \Requester, a simple yet effective scheduler that decides which parts of the video each mobile device should download, adaptive to their cellular rates; \Distributor, an efficient all-to-all dissemination scheme that exploits WiFi overhearing and network coding; and \Broadcast, a networking module that provide pseudo-broadcast capability over WiFi. To the best of our knowledge, \Broadcast is the first networking module to fully exploit the potential of wireless broadcast on Android systems. 

We evaluate the proposed system through both simulations of the NUM solution and experimentation by implementing \MC on a testbed consisting of a number of Android phones. The evaluation results demonstrate that there are significant performance benefits (in terms of decreased download time and increased per-user download rate) compared to alternative approaches, without significant battery cost. A video demonstration and supporting materials can be found on the project website \cite{projectWiki}.

The structure of the rest of the paper is as follows. In Section~\ref{sec:related}, we review related work. In Section~\ref{sec:system}, we present the problem setup and the system overview. In Section~\ref{sec:optimization}, we provide the NUM formulation, solution, and interpretation. In Section~\ref{sec:microcast}, we present the design and implementation of \MC. In Section~\ref{sec:evaluation}, we provide the performance evaluation. Section~\ref{sec:conclusion} concludes the paper.

\section{\label{sec:related} Related Work}
This work combines ideas from network coding, network utility maximization, and cooperation for video streaming. In this section, we discuss the most relevant literature from these areas.


{\flushleft \bf Cooperative Mobile/Wireless Systems.}\quad
When several users are interested in the same content, and they are in proximity of each other, they may be able to use device-to-device connections, \eg WiFi or Bluetooth, to get the content in a cooperative and/or opportunistic way. Opportunistic device-to-device communication is often used for the purpose of offloading the cellular network. For instance, \cite{Ioannidis5}, \cite{HanOffloading14}, and \cite{Whitbeck6} consider a scenario in which device-to-device and cellular connections are used to disseminate the content, considering the social ties and geographical proximity. Instead of offloading cellular networks, our goal is to use cellular and local connectivity so as to  allow each user to enjoy the aggregate downlink rate.

Cooperation among mobile devices for content dissemination or in delay tolerant networking, possibly taking into account social ties, has been studied extensively  \cite{bubbleRAP, hibop}. However, dissemination of content stored on a mobile device is only a special case of our framework, which uses only the local links, but not the cellular downlinks. More importantly, we focus on and exploit single-hop broadcast transmissions, as opposed to multi-hop  communication that exploits mobility (at the expense of delay, which is crucial in our setting) but ignores broadcast.

The idea of using multiple interfaces of mobile devices has been explored before but not in the same way as in this work. For example, \cite{TsaoWifiAggregately} exploits cellular and WiFi interfaces simultaneously to create multiple paths to mobile devices. \cite{SoroushWifiConcurrently} uses concurrent WiFi connections from multiple WiFi hot-spots. \cite{mar} and \cite{diversity} exploit the diversity of multiple interfaces on the same device to achieve better connectivity. In contrast, we use the cellular connections of multiple mobile devices to improve the download rate and jointly utilize the local connections. \cite{train} and \cite{combine} address the same problem using a similar approach but only use unicast communication among peers, as opposed to broadcast used in our work. Thanks to broadcast, even when the local WiFi is congested, and cooperation via unicast is not possible, our scheme may still be able to exploit the benefit of cooperation.

{\flushleft \bf Network Coding in Cooperative, Wireless, and Peer-to-Peer (P2P) Systems.}\quad
Cellular and WiFi links suffer from packet loss due to noise or congestion.
One possible solution to this problem is to have several devices in a close proximity help each other with retransmissions of lost packets. Network coding is particularly beneficial as it can make each retransmission maximally useful to all nodes.  Rate-distortion optimized network coding for cooperative video system repair in wireless P2P networks is considered in \cite{RDO-NC-P2P}. Wireless video broadcasting with P2P error recovery is proposed in \cite{Bopper}. An efficient scheduling approach with network coding for wireless local repair is introduced in \cite{local_repair}. The work in \cite{aalborgICCWorkshops, PictureViewer} and \cite{aalborg1} proposes systems where there are a base station broadcasting packets and a group of smartphone users helping each other to correct errors. Note that base station broadcasting is not implemented in current cellular systems. In contrast, we consider unicast between a base station and a mobile device. In  \cite{CoopVideoRamadan}, a cooperative video streaming system is implemented over personal digital assistants (PDAs) but without network coding. Compared to prior work \cite{Bopper, local_repair, aalborgICCWorkshops, PictureViewer, aalborg1, Hua_P2P}, where each phone downloads all the data, and the local links are used for {\em error recovery}, our scheme {\em jointly} utilizes two interfaces, \eg 3G and WiFi, for data delivery and uses network coding over WiFi to help local repair and make scheduling easier.

Network coding has also been applied to P2P networks for content distribution \cite{microsoft_work, HowPracticalNC, Lava} and live streaming  \cite{R2, Usee}. An excellent review is presented in \cite{RNCinP2Preview}. We will show that, in the micro setting, our local cooperation scheme, \Distributor, significantly outperforms state-of-the-art P2P schemes, including  the widely used BitTorrent \cite{bittorrent} as well as the network coding-based R2 \cite{R2}. This is because \Distributor  is explicitly optimized for the micro setting: it exploits WiFi overhearing and network coding and avoids unnecessary redundancy.

{\flushleft \bf Network Utility Maximization (NUM) of Coded Systems.}\quad
The NUM framework is useful for understanding how different layers and algorithms, such as, flow control, congestion control, and routing, should be designed and optimized \cite{tutorial_doyle}, \cite{book_srikant}. The NUM framework has been used in the past to design algorithms when network coding is employed. The problem of establishing minimum-cost multicast connections over coded wired and wireless networks is considered in \cite{minimum_cost_multicast} and is extended for end-to-end rate/congestion control over wired coded networks in \cite{opt_multicast_nc}. A cross-layer optimization framework including routing and scheduling to maximize throughput over coded wireless mesh networks  for multicast flows is studied  in \cite{opt_framework_in_wireless}. Linear optimization models for computing a high-bandwidth routing strategy for media multicast in coded wireless networks are proposed in \cite{opt_models_in_wireless}.

The NUM framework has also been applied to P2P networks with network coding. In \cite{minghua_P2P}, the aggregate application-specific utility is maximized by distributed algorithms on peers, which are constrained by their uplink capacities. \cite{Shao_P2P} extends \cite{minghua_P2P} by considering node capacities and  constraints on both node upload capacity and node download capacity. In \cite{Liu_P2P}, performance bounds for minimum server load, maximum streaming rate, and minimum tree depth under different peer selection constraints are derived but without  network coding. Optimal bandwidth sharing in multi-swarm multi-party P2P video-conferencing systems with helpers is considered in \cite{Liang_P2P}. Multi-rate P2P multi-party conferencing applications, where different receivers in the same group can receive videos at different rates using, \eg scalable layered coding, are considered in \cite{Ponec_P2P}. The Implicit-Primal-Dual scheme for flow control in live streaming P2P systems is introduced in \cite{Tomozei_P2P}. \cite{Hua_P2P}, which is the closest to our work, proposes a scalable video broadcast/multicast  scheme that efficiently integrates scalable video coding, 3G broadcast, and ad-hoc forwarding so as to balance the system-wide and worst-case video quality of all viewers at 3G cell.

The differences between our work and previous work, at the intersection of NUM and network coding, are the following: (i) our work utilizes multiple cellular unicast links, which is the case in practice, while in prior work, \eg \cite{Hua_P2P}, multicast is assumed deployed on the cellular link; (ii) the NUM formulation in our work takes into account both the cellular links and WiFi transmissions; and (iii) our work exploits network coding and broadcast over WiFi links.

{\flushleft \bf Network Coding in Practice \& Its Implementation.}\quad Network coding has been implemented before on  WiFi testbeds. For example, COPE \cite{cope} is a well-known practical scheme for one-hop network coding across unicast sessions in wireless mesh networks. \cite{IPTV-VTC} proposes a cooperative IPTV system with pseudo-broadcast to improve reliability. Medusa \cite{Medusa} considers a scheme in which multiple unicast flows (video streams) are transmitted from a base station to clients with network coding. This scheme considers rate adaptation and video packet scheduling jointly. SenseCode \cite{sensecode} proposes a collection protocols, that utilize network coding and overhearing, for sensor networks.
 The practicality of random network coding over iPhones is discussed in \cite{RNC-iPHone}. A toolkit to make network coding practical for system devices, from servers to smartphones, is introduced in \cite{Tenor}. \cite{aalborgICCWorkshops} implements network coding on mobile devices and presents the performance in terms of throughput, delay, and energy consumption. \cite{PictureViewer} extends \cite{aalborgICCWorkshops} for picture transmission. In \cite{GestureFlow}, a gesture broadcast protocol is designed for concurrent gesture streams in multiple broadcast sessions over smartphones using inter-session network coding.

To the best of our knowledge,  our work is the first  to implement  network coding and overhearing on the Android platform.  These tasks are much more challenging on Android devices than on laptops, as explained in Section~\ref{sec:microcast}. Network coding has been implemented before on other types of phones \cite{aalborgICCWorkshops, PictureViewer, RNC-iPHone} as well; however, it has not been combined with overhearing, which is one of the key ingredients of our system.


{\flushleft \bf Our work in perspective.}\quad
The particular combination of the three ingredients: cooperation, network coding, and (pseudo)broadcast on Android-based mobile devices, is optimized in the micro setting and differentiates this work from previous schemes.   The contributions of this work lie in (i) the NUM formulation and solution of the problem, and (ii) the practical design and implementation of \MC guided by the theoretical analysis. The theory and systems parts were previously presented in \cite{thisAllerton} and \cite{thisMobisys}, respectively. This journal paper combines these two parts and explains their interaction. For example, it explains how the solution of the NUM framework guides the design of \MC, and conversely, how the performance of \MC validates the NUM framework's solution. Along the way, we showcase key design decisions made to overcome challenges when translating theoretical results into practical systems.

\section{\label{sec:system}System Overview}

We consider the scenario presented in Fig.~\ref{fig:MC-setting-model} (a): a group of mobile device users, within proximity of each other,  are interested in downloading  and watching the same video at roughly the same time. We use cooperation among the mobile devices, where each device simultaneously uses two  interfaces: the cellular interface to connect to the server, and the local interface (WiFi) to  connect to all other users in the group. We can use the two connections (cellular and WiFi) on each device simultaneously and independently, as discussed in Section~\ref{subsec:2-interfaces}. Note that cellular and WiFi connections are used in parallel, but one connects directly to the server (via cellular), while the other connects to the other mobile devices (via WiFi); we do not use both connections to connect a user directly to the server through two different paths. Each mobile device downloads segments of the video from the server and shares these segments with the rest of the group locally. The base stations may be the same or different for different users, depending on their location and the service providers they use.

{\flushleft \bf Source and Flows.}\quad In our analysis, we consider the system model presented in Fig.~\ref{fig:MC-setting-model} (b): the source transmits a video to a group of mobile devices. This model is a simplified version of Fig.~\ref{fig:MC-setting-model} (a) in that the {\em source} includes the video source, video proxies, and base stations.  This allows us to focus on the bottlenecks of the system, namely the cellular and the local area WiFi links. The links between the source and proxies and the links between a proxy and the base stations are typically high capacity links, thus they are not the bottleneck of our system.

Let $\Nset$ be the set of mobile devices (nodes) in the system. The source transmits a video flow of rate $x$ to the nodes. The video flow is associated with a utility function, $U(x)$, which we assume to be a strictly concave function of $x$. 

{\flushleft \bf Cellular Setup.}\quad Each node $i \in \Nset$ is connected to the video source via a cellular link. The cellular link rate is $C_i$ and the loss probability is $p_i$.  We consider $|\Nset|$ parallel interference-free links\footnote{Since the nodes may connect to different base stations and the interference of cellular links are handled by the base stations, we assume that the downlinks are interference free from our perspective.} connecting the source to each node. In practice, some mobile devices, \eg node $i$, may not have a cellular connection. Our system model and analysis capture this case by setting $C_i=0$. We denote the part of video flow transmitted over a cellular link towards node $i$ to help node $j$ by $x_{i,j}$, and $x_{i,i}$ is the flow rate over the cellular link towards node $i$ for its own usage.

{\flushleft \bf WiFi Setup.}\quad Each node $i \in \Nset$ is connected to other nodes in the local area through WiFi. The capacity between nodes $i$ and $j$ is $C_{i,j}$, and the loss probability is $p_{i,j}$. We consider the interference model in \cite{gupta_interference_model}: each node can either transmit or receive through WiFi, and all transmissions in the range of the receiver are considered interfering. As we consider a group of nodes within proximity of each other and do not consider multi-hop packet transmissions, any transmission in the local area interferes with any other transmission, and only one node can transmit at a time.

We consider {\em pseudo-broadcast} transmissions over WiFi, where a node $i$ transmits packets to a node $j$ using unicast transmissions, and each neighboring node overhears and makes use of the overheard packets. Consider a set of nodes $\Jset$, $\Jset \subseteq \Nset$. The pseudo-broadcast rate from node $i$ to all nodes in $\Jset$, is $f_{i,\Jset}$. Note that the set of links $i$--$j$, $j \in \Jset$, is called a hyperlink. The transmission rate over link $i$--$j$, \ie from node $i$ to node $j$, is $g_{i,j}$. The relationship between $g_{i,j}$ and $f_{i,\Jset}$ is clarified in Section~\ref{sec:optimization}.

{\flushleft \bf Loss Model.}\quad In our formulation and analysis, we assume that $p_i$ and $p_{i,j}$ are independent and identically distributed loss probabilities. In practice, the channel model may follow a different (and most probably non-independent and non-identical) distribution. Our formulations could be extended to include more general distributions, and our system implementation (described in Section \ref{sec:microcast}) does not require the knowledge of loss probabilities and probability distributions.

{\flushleft \bf Network Coding.}\quad We consider network coding employed at each mobile device and not at the video source. Each mobile device receives video from the source through its cellular link, performs network coding, and transmit network coded video packets to other mobile devices through WiFi. In our analysis, we consider that the size of the video file is sufficiently large, \ie there always exist packets for transmission. Network coding is performed over the large video file: each coded packet is a linear combination of all packets in the video file. However, in the implementation, we use the practical generation-based network coding \cite{practical_NC}.  Implementation details are provided in Section~\ref{subsec:micro-ncp2}.

{\flushleft \bf  Cooperation.}\quad Several nodes interested in the same video form a single cooperating group\footnote{ In general, the nodes can join or and leave the group according to some rules. However, in this paper, we consider that all the nodes cooperate to form a single group. }. The users in a group  know and trust each other, as it is the case in the motivating examples we provided in the introduction.
We consider two transmission policies in the local area: {\em pseudo-broadcast} and {\em unicast}. (Note that our system implementation in Section~\ref{sec:microcast} corresponds to the pseudo-broadcast policy, and our baseline implementation in Section~\ref{sec:microcast} corresponds to the unicast policy.) In our setup, each node $i$ maintains one input queue and several output queues each which corresponds to a neighboring node. In both policies, each node $i$ receives packets from the source or from its neighbors, stores them in its input queue, and decodes them. The packets received from the source are also put in the output queues. When a transmission opportunity arises (either using NUM scheduling policy or a standard MAC protocol such as 802.11), node $i$ transmits a (possibly coded) packet from an output queue to the corresponding node.

\medskip

Section \ref{sec:optimization} provides the analysis of the system described above, following a network utility maximization framework. We formulate the problem, provide a distributed solution, and interpret structural properties of this optimal solution. In Section \ref{sec:microcast}, we use the insight from the analysis to design a system with the components and algorithms mimicking the optimal solution. We implement it on an Android platform, and we evaluate its performance in \ref{sec:evaluation}.  

\section{\label{sec:optimization} Network Utility Maximization}
As described in Section~\ref{sec:system}, the source transmits video with rate $x$. For node $i \in \Nset$, we consider $N$ different rates: $x_{i,1}, x_{i,2}, \cdots, x_{i,N}$, where the rate $x_{i,j}, j \in \Nset$, is the rate of data transmitted from the source to node $i$ to help node $j$. Our goal is to maximize the utility $U(x)$, which is a strictly concave function of the video source rate $x$. In the rest of this section, we use a network utility maximization (NUM) framework to formulate this problem for the two cooperation policies: pseudo-broadcast and unicast transmissions.

\subsection{Formulation}
{\flushleft \bf Cooperation Policy: Pseudo-Broadcast.}\quad
In this policy, we consider the case that pseudo-broadcast (unicast + overhearing) is available in the local area. If network coding is used in the local area, the NUM problem is formulated as follows:
\begin{align} \label{opt:eq1}
\mbox{\bf P1: }\max_{x} \mbox{ } & U(x) \nonumber \\
\mbox{such that} \mbox{\quad}  & \mbox{1.\quad} \sum_{i \in \Nset} x_{i,j} - x   \geq 0, \mbox{ } \forall j \in \Nset \nonumber \\
& \mbox{2.\quad} g_{i,j} - x_{i,j} \geq 0, \mbox{ } \forall i \in \Nset, j \in \Nset \setminus \{i\} \nonumber \\
& \mbox{3.\quad} x_{i,j} \leq C_{i}(1 - p_{i}), \mbox{ } \forall i \in \Nset, j \in \Nset \nonumber \\
& \mbox{4.\quad} g_{i,j} \leq \sum_{\Jset|j \in \Jset} f_{i,\Jset}, \mbox{ } \forall i \in \Nset, j \in \Nset \setminus \{i\} \nonumber \\
& \mbox{5.\quad} f_{i,\Jset} \leq \min_{j \in \Jset}\{C_{i,j}(1 - p_{i,j})\} \, \tau_{i,\Jset}, \mbox{ } \forall i \in \Nset, J \in \Hset \nonumber \\
& \mbox{6.\quad} \sum_{i \in \Nset} \sum_{J \in \Hset} \tau_{i,\Jset} \leq \gamma
\end{align}
The first constraint is the flow conservation constraint at the source. It requires that the total received rate by node $j$, $\sum_{i \in \Nset} x_{i,j}$, be larger than the targeted video rate, $x$. The second constraint is the flow conservation at the mobile devices. It requires that the outgoing  rate from node $i$ to $j$, $g_{i,j}$, should be larger than or equal to incoming rate, $x_{i,j}$. The third constraint is the capacity constraint in the downlink. Note that the third constraint is equivalent to $\max_{j \in \Nset} \{ x_{i,j} \} \leq C_{i}(1 - p_{i})$, $\forall i \in \Nset$. This constraint is sufficient to represent the capacity constraint in the downlink because the content of flows to help different nodes does not need to be different, \ie the information content of $x_{i,j}$ and $x_{i,k}$, $j \in \Nset$, $k \in \Nset$, could be the same.

The fourth constraint relates the per link transmission rates, $g_{i,j}$, and the pseudo-broadcast rate, $f_{i,J}$. This constraint requires that $g_{i,j}$ should be less than the total of flow rates over all hyperarcs $J$ that lead from $i$ to $j$. This relationship reflects the fact that we employ pseudo-broadcast.

The fifth constraint is the capacity constraint in the local area. In this constraint, $\tau_{i,\Jset}$ is the percentage of time that the hyperlink $\{i ,\Jset\}$ is used. The pseudo-broadcast rate, $f_{i,\Jset}$, should be less than the minimum available capacity from node $i$ to any node $j$, $j \in \Jset$. This constraint reflects the fact that network coding is used in the local area. Network coding helps to improve the broadcast transmission rate. In particular, if network coding is not employed, the fifth constraint should be
\begin{align} \label{opt:noNC_constraint}
f_{i,\Jset} \leq \min_{j \in \Jset}\{C_{i,j}\} \,\cdot\, \prod_{j \in \Jset} (1 - p_{i,j}) \,\cdot\, \tau_{i,\Jset}, \mbox{ } \forall i \in \Nset, J \in \Hset\,.
\end{align}
The reason is that a pseudo-broadcast transmission is only successful if it is successful over all links from $i$ to $j$, $j \in \Jset$, when network coding is not used. This is why the product term is used. We refer to this cooperation policy as {\em Pseudo-Broadcast \& No-NC}. In contrast, when network coding is used, each transmission is beneficial to any node that receives it correctly, independently of other transmissions. Thus, the capacity constraint is improved to $\min_{j \in \Jset}\{C_{i,j}(1 - p_{i,j})\}$.

The last constraint captures  time sharing: time sharing parameters, $\tau_{i,\Jset}$, should be summed up to a provisioning factor, $\gamma$, where $\gamma \leq 1$.

{\flushleft \bf Cooperation Policy: Unicast.}\quad
In this policy, we consider the case that unicast transmissions are used in the local area (WiFi), and network coding is not employed. The NUM problem is formulated as follows:
\begin{align} \label{opt:eq2}
\mbox{\bf P2: }\max_{x} \mbox{ } & U(x) \nonumber \\
\mbox{such that} \mbox{\quad}  & \mbox{1.\quad} \sum_{i \in \Nset} x_{i,j} - x \geq 0, \mbox{ } \forall j \in \Nset \nonumber \\
& \mbox{2.\quad} g_{i,j} - x_{i,j} \geq 0, \mbox{ } \forall i \in \Nset, j \in \Nset \setminus \{i\} \nonumber \\
& \mbox{3.\quad} x_{i,j} \leq C_{i}(1 - p_{i}), \mbox{ } \forall i \in \Nset, j \in \Nset \nonumber \\
& \mbox{4.\quad} g_{i,j} \leq C_{i,j}(1 - p_{i,j})\tau_{i,j}, \mbox{ } \forall i \in \Nset, j \in \Nset \setminus \{i\} \nonumber \\
& \mbox{5.\quad} \sum_{i \in \Nset} \sum_{j \in \Nset \setminus \{i\}} \tau_{i,j} \leq \gamma
\end{align} Note that the first three constraints of Eq.~(\ref{opt:eq2}) are the same as Eq.~(\ref{opt:eq1}). The fourth constraint is the capacity constraint in the local area: the transmission rate from node $i$ to $j$, $g_{i,j}$, should be less than the capacity of the link and the percentage of time the link is used for that transmission, $\tau_{i,j}$. The last constraint is the time sharing constraint similar to Eq.~(\ref{opt:eq1}).

\subsection{Solutions}
{\flushleft \bf Solution for P1.}\quad
Let us first consider the solution for P1 in Eq.~(\ref{opt:eq1}). By relaxing the first and second constraints in Eq.~(\ref{opt:eq1}) via Lagrangian relaxation, we have the following Lagrangian function:
\begin{align} \label{relax:eq1}
& L(x,\lambda,\eta) = U(x)+ \sum_{j \in \Nset} \lambda_{j} (\sum_{i \in \Nset} x_{i,j} - x)  + \sum_{i \in \Nset} \sum_{j \in \Nset \setminus \{i\}} \eta_{i,j} (g_{i,j} - x_{i,j})\,,
\end{align} where $\lambda_{j}$ and $\eta_{i,j}$ are Lagrange multipliers. Eq.~(\ref{relax:eq1}) is expressed as
\begin{align} \label{relax:eq2}
L(x,\lambda,\eta) &= U(x)+ \sum_{i \in \Nset} \sum_{j \in \Nset} \lambda_{j} x_{i,j} - x \sum_{j \in \Nset} \lambda_{j} + \sum_{i \in \Nset} \sum_{j \in \Nset \setminus \{i\}} \eta_{i,j} g_{i,j} - \sum_{i \in \Nset} \sum_{j \in \Nset \setminus \{i\}} \eta_{i,j} x_{i,j} \nonumber \\
~ &= U(x) - x \sum_{j \in \Nset} \lambda_{j} + \sum_{i \in \Nset} \sum_{j \in \Nset} x_{i,j} (\lambda_{j} - \eta_{i,j})  + \sum_{i \in \Nset} \sum_{j \in \Nset} \eta_{i,j}  g_{i,j}\,.
\end{align}
The Lagrangian function in Eq.~(\ref{relax:eq2}) is decomposed into several sub-problems, each of which solves the optimization problem for one variable. We provide the decomposed solution in the following.

{\em Flow Control at the Source:} First, we solve the Lagrangian function with respect to $x$:
\begin{align} \label{eq:sourceratecontrol}
x = (U')^{-1}(\sum_{j \in \Nset} \lambda_{j})\,,
\end{align} where $(U')^{-1}$ is the inverse function of the derivative of $U$. This part of the solution is interpreted as the flow control at the source. Note that $\lambda_{j}$ is the Lagrangian multiplier, and it can be interpreted as the queue size at the source for packets to be transmitted to node $j$. (The reason for this interpretation will be provided later in this section.) By considering $\lambda_{j}$ as the queue size and taking into account that $U(x)$ is a strictly concave function of $x$, it could be observed from Eq.~(\ref{eq:sourceratecontrol}) that $x$ is inversely proportional to the sum of the queues for all nodes in a cooperating group. This policy regulates the amount of traffic that are generated at the source and inserted into the output queues at the source based on the number of packets in the queues to avoid congestion (or buffer overflows). In our implementation, the video source, \eg YouTube server, has its own algorithms to regulate the generated traffic. Therefore, we do not use this part of the solution in our implementation and rely on the flow control mechanism of the video source (\eg TCP) to avoid buffer overflows.

{\em Downlink Rate Control:} Second, we solve the Lagrangian function with respect to $x_{i,j}$:
\begin{align} \label{eq:downlinkratecontrol}
\max_{\mathbf x} \mbox{ } & \sum_{i \in \Nset} \sum_{j \in \Nset} x_{i,j}(\lambda_{j} - \eta_{i,j}) \nonumber \\
\mbox{s.t.} \mbox{ }  & x_{i,j} \leq C_{i} (1 - p_{i}), \mbox{ } \forall i \in \Nset, j \in \Nset\,
\end{align}
where $\mathbf x$ is the vector of $\{x_{i,j}\}_{i,j \in \Nset}$. Here, the Lagrange multipliers $\eta_{i,j}$ can be considered as the size of the queue at node $i$ for packets that should be transmitted from node $i$ to node $j$. (The reason for this interpretation will be provided subsequently.) According to Eq.~(\ref{eq:downlinkratecontrol}), the transmission rate, $x_{i,j}$, should be set to zero if the difference of the queue size at the source for node $j$, $\lambda_{j}$, is less than the queue size at node $i$ for node $j$, $\eta_{i,j}$.
In our implementation, we do not have control at the source, and information about $\lambda_{j}$ is not available. Nevertheless, if we assume that $\lambda_{j}$ is fixed for all $j$ (which is a valid assumption, for example, for TCP flows in larger time scale as compared to TCP's own clock) then $x_{i,j}$ becomes inversely proportional to the queue size, $\eta_{i,j}$, at node $i$. Therefore, if $\eta_{i,j}$ is large,  $x_{i,j}$ should be small and vice versa. Based on this observation, and also not to overload queues at the mobile devices, our practical download algorithm implemented in \Requester requests more packets for the mobile devices with small queue sizes, and requests less packets for the devices with large queue sizes. The details of the algorithm are presented in Section~\ref{subsec:micro-download}.

{\em Local Area Rate Control and Scheduling:}
Third, we solve the Lagrangian function for $\tau_{i,\Jset}$. It could be observed from the fourth and fifth constraints of Eq.~(\ref{opt:eq1}) that the optimal value of $g_{i,j}$ is $\sum_{\Jset|j \in \Jset} \min_{j \in \Jset}\{C_{i,j}(1-p_{i,j})\}\tau_{i,\Jset}$. Therefore, the local area rate control and scheduling problem could be expressed as follows:
\begin{align} \label{eq:localarearate_P1}
\max_{\mathbf \tau} \mbox{ } & \sum_{i \in \Nset} \sum_{\Jset \in \Hset} \tau_{i,\Jset} (\sum _{j \in \Jset} \eta_{i,j} \min_{j \in \Jset}\{C_{i,j}(1-p_{i,j})\} ) \nonumber \\
\mbox{s.t.} \mbox{ }  &  \sum_{i \in \Nset} \sum_{\Jset \in \Hset} \tau_{i,\Jset} \leq \gamma\,,
\end{align}
where $\mathbf \tau$ is the vector of $\{\tau_{i,\Jset}\}_{i \in \Nset, \Jset \subseteq \Nset}$. Eq.~(\ref{eq:localarearate_P1}) determines the percentage of time that a hyperarc $\{i,\Jset\}$ is used for transmitting packets. It could be observed from  Eq.~(\ref{eq:localarearate_P1}) that the pseudo-broadcast link $\{i,\Jset\}$ is activated if the $\sum _{j \in \Jset} \eta_{i,j} \min_{j \in \Jset}\{C_{i,j}(1-p_{i,j})\} $ term is larger than that of the other links. This means that the pseudo-broadcast link, which could transmit more packets, \ie the pseudo-broadcast link with the largest  $\sum _{j \in \Jset} \eta_{i,j}$, and better capacity, is activated for transmission. Our practical implementation is based on this idea. In particular, the node having a segment needed by the largest number of devices, which (broadly) corresponds to $\sum _{j \in \Jset} \eta_{i,j}$, takes the opportunity and transmits the segment. Specifically, the \Distributor algorithm pseudo-broadcasts the segment that is needed by the largest number of devices immediately. More details of \Distributor are provided in Section~\ref{subsec:micro-ncp2}.

{\em Queue Update:}
The decomposed parts of the Lagrangian, \ie Eqs.~(\ref{eq:sourceratecontrol}), (\ref{eq:downlinkratecontrol}), (\ref{eq:localarearate_P1}) and the Lagrange multipliers  $\lambda_{j}$ and $\eta_{i,j}$ can be solved iteratively via sub-gradient algorithms. In particular, $\lambda_{j}$ can be iteratively solved as follows:
\begin{align} \label{eq:queueupdatesource}
\lambda_{j}(t+1) = \{ \lambda_{j}(t) + \beta_{t} [x(t) - \sum_{i \in \Nset} x_{i,j}(t) ] \}^{+}, \mbox{ } \forall j \in \Nset\,,
\end{align} where $t$ is the iteration number, $\beta_{t}$ is a small constant (step size of the gradient descent algorithm), and the $\{\}^+$ operator makes the Lagrange multipliers positive, \ie $\{a\}^{+}=\max\{a,0\}$. The Lagrange multiplier $\lambda_{j}$ is interpreted as the queue size at the source for the packets to be transmitted to node $j$ because it is updated as the difference between the incoming traffic, $x$, and the outgoing traffic, $\sum_{i \in \Nset} x_{i,j}$. 

Similarly, $\eta_{i,j}$ can be solved iteratively as follows:
\begin{align} \label{eq:queueupdatelocal}
& \eta_{i,j}(t+1) = \{\eta_{i,j} (t) + \beta_{t} [x_{i,j}(t) -  g_{i,j}(t)] \}^{+},  \mbox{ } \forall i \in \Nset, j \in \Nset\,.
\end{align} The Lagrange multiplier $\eta_{i,j}$ is interpreted as the queue size at node $i$ for packets to be transmitted from node $i$ to node $j$ because it is updated as the difference between the incoming rate, $x_{i,j}$, and the outgoing rate, $g_{i,j}$. In the interest of space, we provide the convergence analysis of the solution in \cite{thisTechRep}.

{\flushleft \bf Solution for P2.}\quad
Now, let us consider the solution for P2 in Eq.~(\ref{opt:eq2}). The decomposed solution of P2 exactly follows Eq.~(\ref{eq:sourceratecontrol}) for the flow control at the source, Eq.~(\ref{eq:downlinkratecontrol}) for the downlink rate control, and Eq.~(\ref{eq:queueupdatesource}) and Eq.~(\ref{eq:queueupdatelocal}) for the queue updates at the source and local nodes, respectively. The only different part is the local area rate control. Note that the optimal value of $g_{i,j}$ is $g_{i,j} = C_{i,j}(1 - p_{i,j})\tau_{i,j}$. The rate control for Eq.~(\ref{opt:eq2}) is as follows:
\begin{align} \label{eq:localarearate_P2}
\max_{\bf \tau} \mbox{ } & \sum_{i \in \Nset} \sum_{j \in \Nset} \eta_{i,j} C_{i,j} (1 - p_{i,j})\tau_{i,j} \nonumber \\
\mbox{s.t.} \mbox{ }  &  \sum_{i \in \Nset} \sum_{j \in \Nset} \tau_{i,j} \leq \gamma
\end{align} Similar to Eq.~(\ref{eq:localarearate_P1}), Eq.~(\ref{eq:localarearate_P2}) determines the percentage of time that a link $(i,j)$ is used for transmitting packets.

\section{MicroCast: Design and Implementation}\label{sec:microcast}
\label{sec:microcast}

\begin{figure*}[t!]
\centering
  \subfigure[Architecture]{\includegraphics[width=0.32\textwidth]{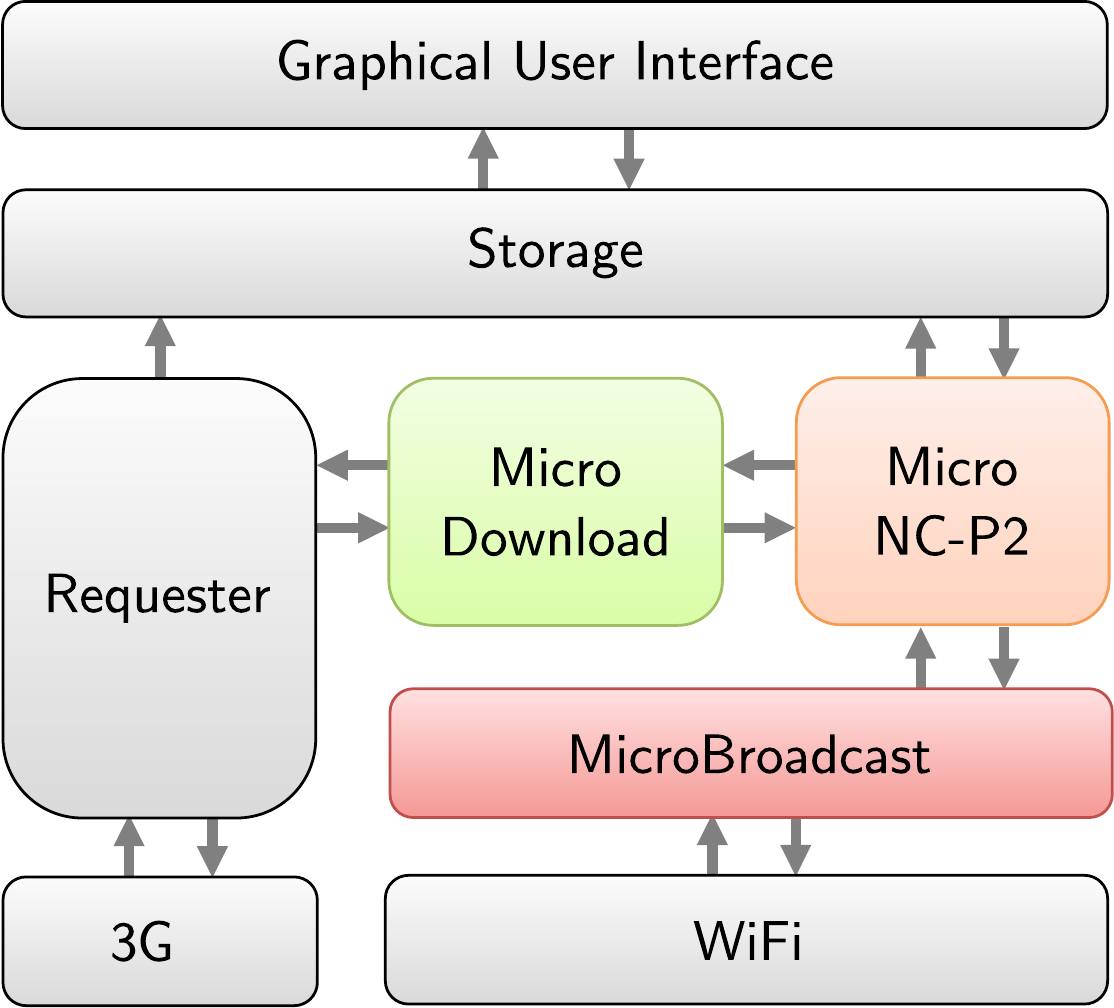}}\hspace{20pt}
  \subfigure[Distributors]{\includegraphics[width=0.17\textwidth]{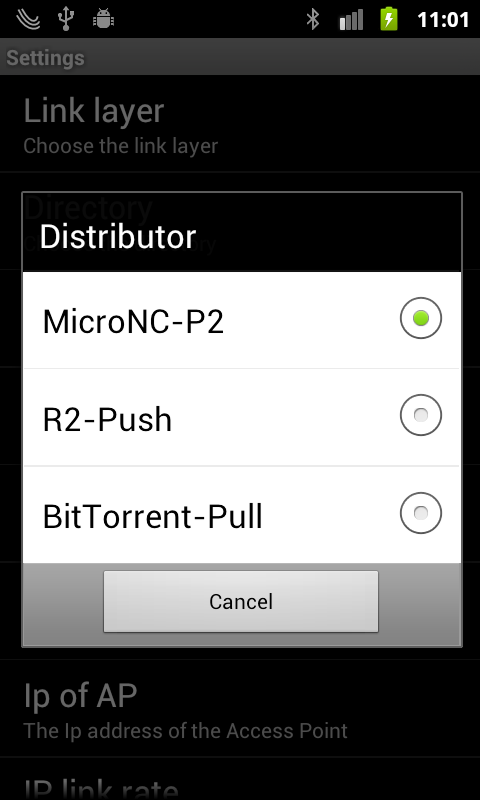}}\hspace{10pt}
  \subfigure[Downloading]{\includegraphics[width=0.17\textwidth]{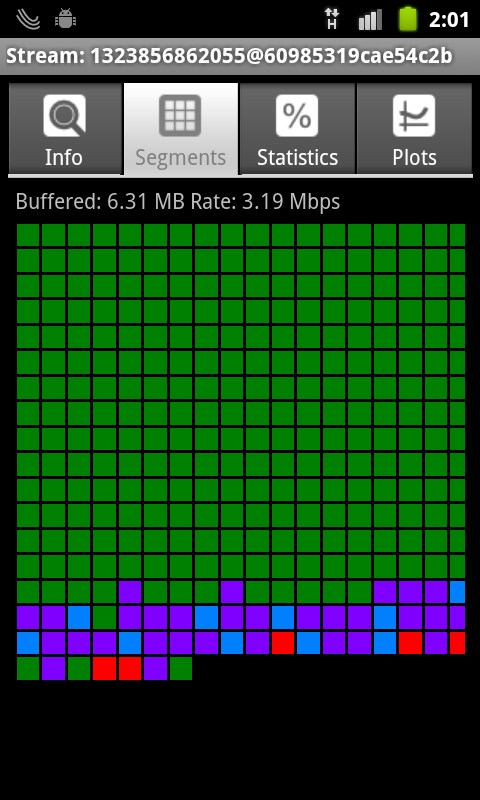}}\hspace{10pt}
  \subfigure[Statistics]{\includegraphics[width=0.17\textwidth]{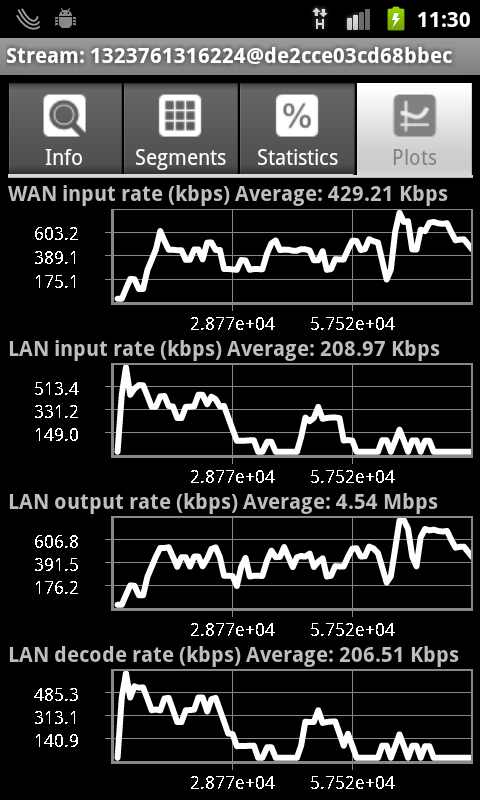}}
\caption{The MicroCast Architecture and snapshots from the Graphical User Interface. {\bf (a)} Architecture and main components of Microcast. {\bf (b)} Choosing a scheme for local cooperation. {\bf (c)} Visualization of downloading: packets already stored (green), currently downloaded from the cellular (blue), from neighbors through WiFi (purple) and missing (red). {\bf (d)} Measurements of download rates over different links. \label{fig:microCast-arc-gui}}
\end{figure*}

In this section, we present \MC, the prototype system that we develop on the Android platform based on the insight gained from the structural properties of the optimal NUM solution. We developed  \MC mostly in Java with some parts in C. It currently  runs on Android 2.3 and 4.0.  Fig.~\ref{fig:microCast-arc-gui} (a) shows the overall architecture of \MC and  the main components: Requester, \Distributor, and \Broadcast. Additional components include Requester, Storage, and Graphical User Interface. In the following, we describe each component in detail.  

\subsection{\Requester}
\label{subsec:micro-download}

\begin{algorithm}[t]
\begin{algorithmic}[1]
\WHILE {there are segments to assign}
  \STATE Find the device with the smallest backlog
  \IF{the backlog of the device is smaller than $K$}
   	  \STATE Schedule the device to download the next segment
  \ELSE
  	  \STATE Sleep until new feedback is received
  \ENDIF
\IF{feedback from device indicates a failure}
    \STATE{Schedule the device to download another segment}
    \STATE{Add the segment that failed to the list of segments}
\ENDIF
\ENDWHILE
\end{algorithmic}
\caption{MicroDownload Algorithm}
\label{scheduler}
\end{algorithm}

This component is present on all devices in the group but runs only on the one that initiates the download. It instructs the requesters of all devices which segments of the video to download from the server. (Recall that a video is divided into multiple segments.) The key idea of \Requester is to assign the next segment to be downloaded to a device which has the smallest backlog, where the backlog refers to the set of segments previously assigned to this device. 

 The \Requester algorithm is summarized in Alg.~\ref{scheduler}. \Requester has a list of segments that should be assigned to the devices. Initially, it assigns a fixed number ($K$) of segments to each device. The devices try to download one after another the segments that are assigned to them. If a device downloads a segment successfully, it notifies \Requester. Otherwise, it reports failure. \Requester re-assigns the failed segments based on the backlogs. This mechanism ensures that no segment remains trapped in any device which has a bad cellular connectivity. This mechanism also adapts segment download to the varying rates and conditions of cellular links. For example, if a device has a bad cellular connection, the segments being handled by it will be reassigned to other devices, which hopefully have better connections. However, \Requester will still assign some segments to the device with a bad cellular connection so that it can start downloading immediately as soon as its connection quality improves.

\subsection{\Distributor}
\label{subsec:micro-ncp2}

\algblockdefx[WHEN]{STARTWHEN}{ENDWHEN}%
	[1][]{{\bf when} #1}%
	{{\bf end when}}%

\algrenewcommand{\algorithmiccomment}[1]{// {\em #1}}

\begin{algorithm}[t]
\begin{algorithmic}[1]
\STARTWHEN {a new segment $s$ is received}
\IF {$s$ is received by the requester}
\STATE \COMMENT {initial push, $m$ is the dimension of $s$}
\STATE Send $m$ coded packet of $s$ to a neighbor
\ENDIF
\STATE Add $s$ to the list of segments to be advertised
\ENDWHEN
\Statex
\STARTWHEN {a packet $p$ is received from $A$}
\IF {$p$ is an advertisement or notification containing $s$}
\STATE \COMMENT {subsequent pulls exploit overhearing}
\STATE Request $A$ for the missing dimensions of $s$
\ELSIF {$p$ is a request for $d$ dimensions of $s$}
\STATE Add this request to the request queue
\ELSIF {$p$ is a coded packet of $s$}
\STATE Progressively decode $s$ using $p$
\ENDIF
\ENDWHEN
\Statex
\STARTWHEN {there is a request for $d$ dimensions of $s$ from $A$}
\STATE \COMMENT {pseudo-broadcast}
\IF {there are other similar requests}
\STATE Let $d$ be largest requested dimension
\STATE Remove these requests from the request queue
\ENDIF
\STATE Send $d$ coded packets of $s$ to $A$
\ENDWHEN
\end{algorithmic}
\caption{\Distributor Algorithm}
\label{distribution}
\end{algorithm} 

This component is responsible for distributing segments using the local WiFi network, exploiting the benefits of overhearing and network coding. At a high level, \Distributor takes advantage of pseudo-broadcast, \ie unicast and overhearing, to reduce the number of transmissions. Furthermore, instead of disseminating plain packets, it disseminates random linear combinations of packets (network coded packets) of the same segment. This is to maximize the usefulness  of overheard packets \cite{monograph}. We will provide details on how we implement network coding subsequently. We term our dissemination scheme \Distributor, where P2 refers to an {\em initial Push} and {\em subsequent Pulls}, as explained below.

Our \Distributor is designed based on traditional pull-based P2P dissemination schemes, such as BitTorrent. In \Distributor, a segment $s$ is divided into $m$ plain packets. $m$ is also known as the {\em dimension} of the linear space $s$ or the size of a network coding generation. A device, $A$, periodically advertises the segments that it currently has to its neighbors. Then, a neighbor, say device $B$, requests segments that it does not have based on the advertisement. Upon receiving the request, $A$ sends the requested segments to $B$. More specifically,
\begin{itemize}
\item When $B$ requests a segment $s$ from $A$, it takes into account previously overheard coded packets of segment $s$. In particular, it explicitly indicates in the request how many additional coded packets (missing dimensions) it needs to receive to decode $s$. This reduces the number of coded packets to be sent.
\item When $A$ is about to serve a segment $s$ requested by $B$, it first checks if there are pending requests for the same segments from other neighbors. If there are, it finds the maximum number of coded packets requested among these requests. Denote this maximum number by $d$. If there is none, $d$ is the number of coded packets requested by $B$. Afterwards, $A$ serves $d$ coded packets of segment $s$ to $B$. The other devices, which need up to $d$ coded packets of $s$, should be able to get them through overhearing.
\end{itemize}

\begin{figure*}[t!]
\centering
  \subfigure[Space-Time Diagram of \Distributor]{\label{fig:space-time-NC2P}\includegraphics[height=50mm]{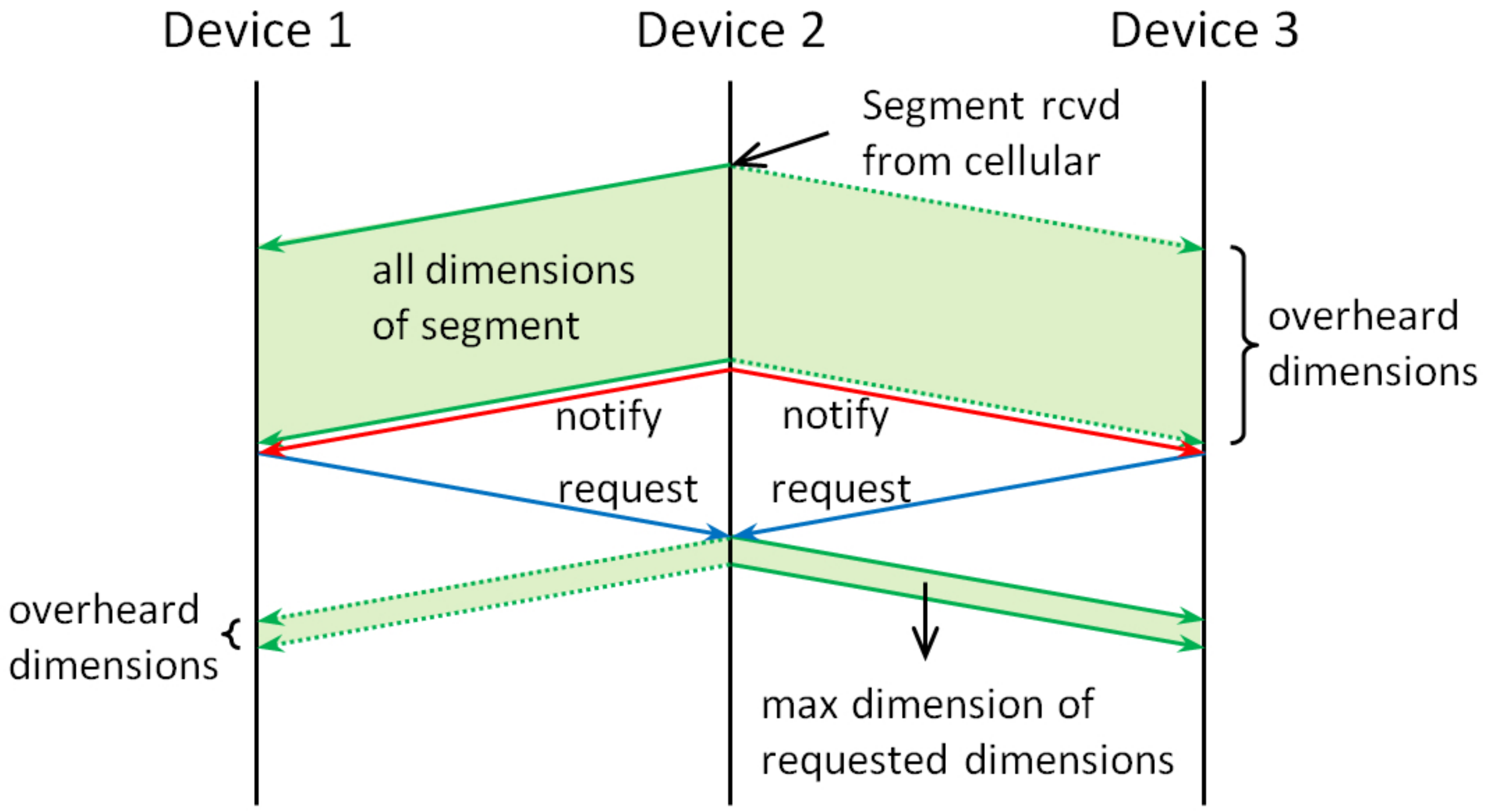}\hspace*{15mm}}
  \subfigure[\Broadcast]{\label{fig:micro-broadcast (and pseudo-adhoc mode)}\includegraphics[height=50mm]{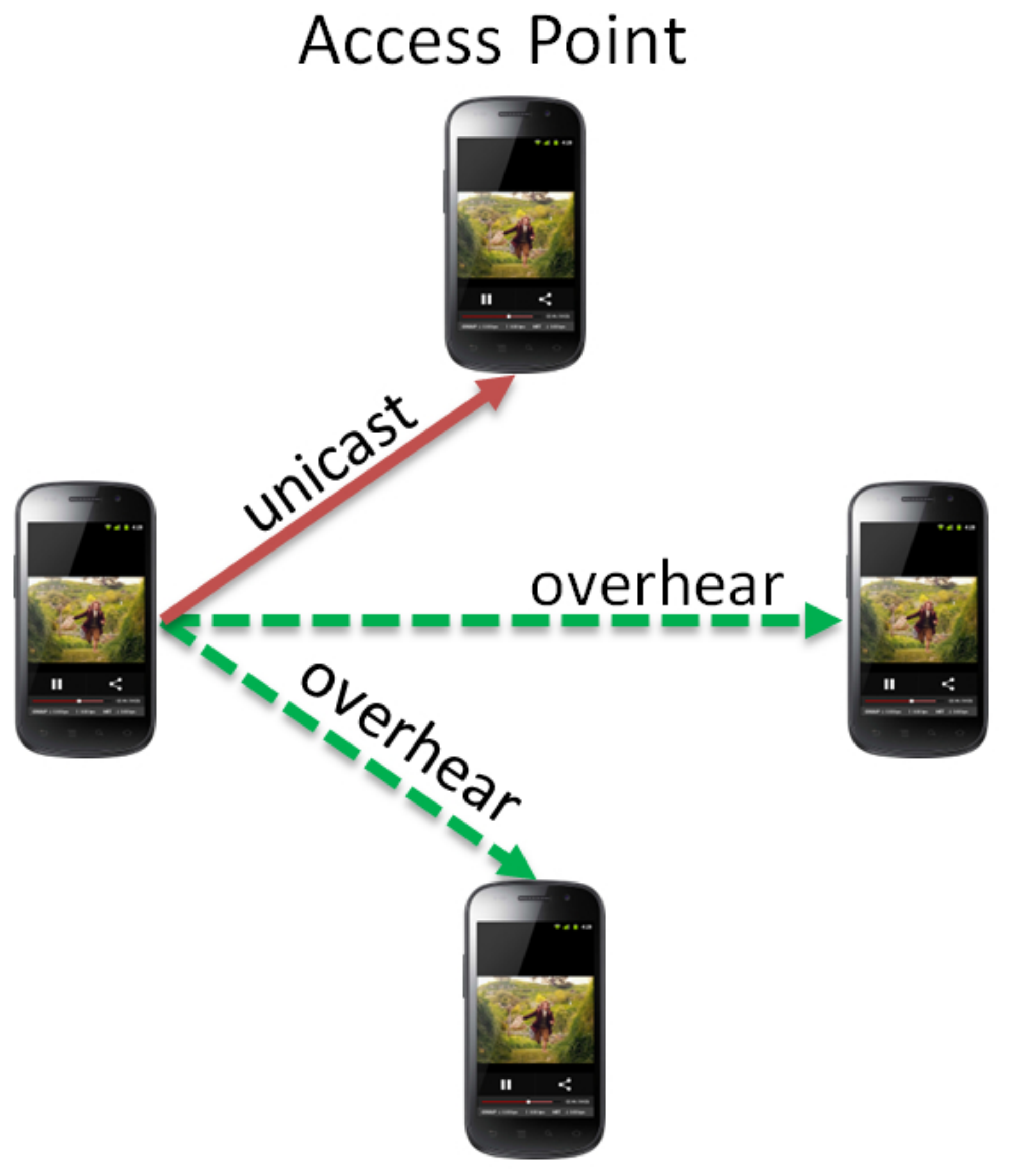}}
\caption{(a) \textbf{\Distributor} is the local dissemination schemes that exploits network coding and pseudo-broadcast provided by {\Broadcast}: Coded packets of a just downloaded video segment are initially pushed by pseudo-broadcasting. Then, the neighbors request for additional coded packets necessary for decoding. Finally, the recovery takes into account all requests to save bandwidth. (b) \textbf{\Broadcast} uses one device as an Access Point. This AP device does not forward packets. A non-AP device pseudo-broadcasts by unicasting to the AP and the rest overhears. {\Broadcast} achieves one transmission per broadcast, which is the most efficient possible.}
\label{fig:MC}
\vspace{-20pt}
\end{figure*}

After serving $B$, $A$ notifies all devices that requested some coded packets of segment $s$. Upon receiving the notification, these devices check if they received all the necessary coded packets to decode $s$. If not, they send requests for additional coded packets. This is necessary because overhearing is not guaranteed for all coded packets sent by $A$ and for all devices. Finally, the scheme gives higher priority to requests that are closer to the playback time when serving them. Overhearing and unicast effectively allow for pseudo-broadcast. The amount of traffic saved by pseudo-broadcasting segment $s$ depends not only on the quality of the overhearing but also on  the number of requests of segments $s$ from other devices that $A$ processes at the time of broadcasting.

To maximize the traffic savings, we propose an initial push of segment $s$. Specifically, when $A$ finishes downloading segment $s$, it sends $m$ coded packets of $s$ to a randomly selected neighbor before advertising the segment. (Note that this just downloaded segment is beneficial to all neighbors.) By doing so, $A$ ensures that the initial dissemination of segment $s$ is taken into account in subsequent requests of segment $s$ (if any) of $A$'s neighbors. This effectively creates a {\em perfect synchronization} of the reception of the initial requests of segment $s$. We provide the pseudocode of \Distributor distribution algorithm in Alg.~\ref{distribution} and the space-time diagram of \Distributor in Fig.~\ref{fig:space-time-NC2P}.

Last but not least, in order to address loss of request and notification packets, which could lead to incomplete segments, \Distributor includes a {\em recovery thread}. This thread periodically re-requests segments that were requested after a certain amount of time but never received.

{\flushleft \bf Implementing Network Coding.}\quad 
We implement the practical generation-based network coding \cite{practical_NC} over the field GF($2^8$). The video file is divided into multiple segments. A segment represents a coding generation. Each segment is broken down into $m$ packets, where $m$ is the generation size. Each packet contains $n$ bytes, and we treat each byte as a symbol in GF($2^8$). We also augment each packet with the $m$ coding coefficients. Coding coefficients of a coded packet is selected uniformly at random from  GF($2^8$). Each packet can be seen as a vector of length $n+m$ symbols from GF($2^8$).

Let $M$ denote the matrix whose rows are $m$ linearly independent packets (of size $n+m$) of the same segment that a device received: $ M = [ E\,|\,C ]$, where $E$ is the data matrix of size $m \times n$ and $C$ is the  coefficient matrix of size $m \times m$. The original packets of the segment can be recovered by finding the inverse of $C$. In particular, $C^{-1} \cdot [E\,|\,C]  = [B\,|\,I]$, where $B$ is the matrix of size $m \times n$, whose rows are the original packets, and $I$ is the $m \times m$ identity matrix. Inverting $C$ takes $\Theta(m^3)$ and multiplying $C^{-1}$ with $[E\,|\,C]$ takes $\Theta(m^2(n+m))$ in terms of finite field multiplication. Thus, the decoding takes $\Theta(m^3 + nm^2)$ in total. Generating $m$ randomly encoded packets can be done by generating a random coefficient matrix $R$ of size $m \times m$ and  multiplying $R$ with $[B\,|\,I]$. Thus, the encoding of a segment also takes $\Theta(m^3 + nm^2)$.

As described above, network coding is a CPU intensive operation. In \MC,  encoding and decoding must be performed efficiently, at a rate matching that of the local network dissemination. Otherwise, CPU risks to become the bottleneck of the video distribution. Therefore, in our implementation, we explored several ways to optimize the coding speed.

The first method to reduce the CPU usage is to limit the size of the coding generation ($m$). The smaller the number of packets in each segment, the smaller the coding complexity. Using smaller segment sizes, however, reduces the diversity of encoded packets, \ie packets are less likely to bring innovative information to their recipients. Second, we seek to optimize our implementation of network coding. In particular, we test two implementation approaches: pure Java and native code. In the first implementation, the encoding and decoding operations are performed by code that runs in the Dalvik virtual machine. In the second approach, the code runs natively on the device CPU and is invoked through the Java Native Interface. The Java implementation has the advantage of being portable across different hardware platforms, but it is less efficient than the native version. In both implementations, we use table lookups to perform finite field multiplication and division, and we use the bit-by-bit XOR operation to perform addition and subtraction.  In Section~\ref{subsec:sys-eval}, we provide the evaluation of both of these implementations.

\subsection{\Broadcast}
\label{subsec:micro-broadcast}

This component implements a comprehensive networking stack, which operates on current wireless technologies, including WiFi 802.11 and RFCOMM Bluetooth. \Broadcast supports unicast, reliable and un-reliable message exchange between the devices over both WiFi and Bluetooth. It also includes multi-hop routing, network-wide flooding, and peer discovery. The most important functionality that \Broadcast provides is the ability to pseudo-broadcast over WiFi. To the best of our knowledge, this is the first system that provides this capability on top of WiFi on Android devices. Depending on the wireless technology used, features of MicroBroadcast are either implemented using a native mechanism or custom developed. For instance, the Bluetooth implementation re-uses the native peer discovery mechanism while WiFi devices run a custom peer discovery protocol.

{\bf Implementing High-Rate WiFi Broadcast.}\quad Although devices within proximity of each other can, in principle, overhear all transmissions, high-rate broadcast was not possible with the existing modes.  The {\em unicast mode of 802.11} does not exploit broadcast:  it (redundantly) transmits the same packets to each receiver separately.  The {\em broadcast mode of 802.11} has its own disadvantages \cite{Medusa}: (i) it lacks a back-off mechanism, which may harm the performance of other flows;  (ii) its transmission rate is limited to the minimum (base rate, 1 Mbps); (iii) finally, unlike laptops, it is not always possible to adapt the broadcast transmission rate on Android devices due to wireless driver and firmware limitations.

A possible solution is to use {\em pseudo-broadcast}, \ie {\em overhearing}, which combines the benefits of unicast and broadcast. Unicast is used as the transmission mode, but the devices overhear all transmissions in their neighborhood. Pseudo-broadcast combines the desirable properties of unicast (high rate, back-off) with overhearing, which makes it attractive. Although it has been implemented in several frameworks \cite{cope, Medusa, dircast}, when implementing pseudo-broadcast, we faced several challenges that are specific to the Android platform.

First, the devices we use do not readily support the {\em promiscuous mode} due to the constraints imposed by the WiFi firmware and drivers. Therefore, we have to update the WiFi drivers of all the Android 2.3 devices we use, and the firmware in some of them. In particular, we update the WiFi driver and firmware by installing CyanogenMod  7 ROM \cite{cyanogenmod} (a custom Android firmware) on the devices after testing various possible firmwares. With the CyanogenMod 7 ROM, promiscuous mode is available, but only in {\em infrastructure mode} (not in ad-hoc mode).

Second, even with the promiscuous mode enabled, Android does not support pseudo-broadcast mode natively, \ie it does not pass the overheard packets up to the application layer. We have to  develop our own {\em overhearing API} for that purpose ``under the hood of Android,''  by developing our own C library and a C binary executable program that runs as a daemon. They also filter out irrelevant overheard packets  (\ie packets which do not belong to video data that the devices are interested in) so as not to overload the CPU.

Third, as we mentioned above, this pseudo-broadcast implementation works only in infrastructure mode but not ad-hoc mode. Using infrastructure mode has a major disadvantage compared to ad-hoc mode: when a device transmits a packet to another device, the packet has to be relayed by the access point, which results in double amount of traffic.

To avoid this disadvantage of infrastructure mode as well as to exploit the benefits of overhearing, we implement a {\em pseudo-ad-hoc mode}, which is shown in Fig.~\ref{fig:MC}(b). In this mode, a device is chosen to act as the access point (AP). In order for a device to transmit packets to other devices, it transmits them to the AP. These transmissions are overheard opportunistically by the targeted devices. When the AP device receives a packet, it {\em does not} forward it (as it would normally do in the infrastructure mode), since the other devices should already have received it via overhearing. In this manner, we are able to enable overhearing (which is only possible in infrastructure mode), while ensuring only one transmission per packet (which is the case in ad-hoc mode), thus the term {\em pseudo-adhoc mode}.

\subsection{Other Components}

{\flushleft \bf Requester} retrieves segments of the video from the video source. It internally uses components called {\em producers} to retrieve the segments. Our current implementation contains producers for three types of sources: HTTP, file, and content. The first one (HTTP) loads segments from an HTTP server using range requests. The second one (file) loads video from locally available files. Finally, the third one (content) retrieves data using the ContentProvider API of Android (\eg videos can be captured with the device camera).

{\flushleft \bf Storage} is used to cache the received segments for successive playback. The segments are stored in the internal flash memory of the device to keep the application memory requirements low. It is possible to access the segments from the storage either using a Java API, as done by the requester and \Distributor, or via an embedded HTTP server that we have developed. This second interface allows us to play the video stream using the native Android media API.

{\flushleft \bf  Graphical User Interface (GUI)} provides an interface to create video streams, share local videos, start/stop downloading video files, join existing video streams, and play/pause videos. In addition to these basic features, the GUI lets users discover and connect to other devices, specify the wireless interface and algorithm that should be used for the local dissemination, and decide whether the device should collaborate for video downloading. In addition, it provides live feedback and detailed statistics during and after experiments. These functionalities of the GUI facilitate field tests. We provide some screen shots of our GUI in Fig.~\ref{fig:microCast-arc-gui} (b, c, d). Finally, the video can be played while \MC is still downloading; therefore, live streaming is supported.

\subsection{Using Multiple Network Interfaces}
\label{subsec:2-interfaces}

Each device needs to use a network interface as downlink (\eg 3G or WiFi) and another interface (\eg WiFi or Bluetooth) for the local cooperation.  For the connection to the server, we choose 3G over WiFi mainly because it provides ubiquitous Internet access. A second reason is that Bluetooth and WiFi share partially overlapping parts of the spectrum and are often implemented in the same chip; meanwhile,  3G is usually implemented on a different chip and uses a different part of the spectrum. This suggests that using WiFi (for the connection to the server) together with Bluetooth (for the local network) may noticeably decrease the transmission rate, which was indeed the case during our initial experiments. 3G is independent from both Bluetooth and WiFi, so the combination of 3G and either WiFi or Bluetooth does not reduce the transmission rate. For the local connection, we use WiFi instead of Bluetooth because it can support a larger number of connections at higher rates, and more importantly, Bluetooth does not support broadcast, which was a necessary ingredient for achieving \MC's benefits.

In practice, the Android connectivity manager imposes additional challenges when 3G is utilized simultaneously with WiFi. In particular, in order to improve the battery life, every time the WiFi interface is activated, Android turns the 3G interface off.  We solve this problem using an undocumented API that forces routing of packets from the remote server through the 3G interface, therefore preventing the interface from being shut down.

\section{Performance Evaluation}
\label{sec:evaluation}

In this section, we provide simulation results of the NUM solution (Section \ref{subsec:num-eval}) and experimental evaluation of the implemented \MC Android  prototype (Section\ref{subsec:sys-eval}).   We find that (i) the results obtained from analysis agree with experimental evaluation and (ii) \MC brings significant performance benefits, in terms of common download rate, compared to state-of-the-art cooperation schemes in the ``micro'' setting, without significant battery penalty.

\subsection{Simulation of NUM Solution}
\label{subsec:num-eval}

\subsubsection{Simulation Setup}

First, we simulate our NUM solutions in Matlab. We consider the topology shown in Fig.~\ref{fig:MC-setting-model} (b). The simulation includes 1000 iterations, and each simulation is repeated for 10 different random seeds. At each iteration $t$, each link (corresponding to cellular or WiFi) is in the ON or OFF state according to the loss probabilities: $p_i$ and $p_{i,j}$. If a link is on OFF state, the transmitted packet is considered lost; otherwise, it is considered as successfully delivered. The utility function employed in our simulations is $U(x)=log(x)$.

We compare the following four schemes: {\em Pseudo-Broadcast} as a solution of P1 in Section~\ref{sec:optimization}; {\em Pseudo-Broadcast \& No-NC} as a solution of P1 without network coding, \ie using Eq.~(\ref{opt:noNC_constraint}) in P1 instead of the fifth constraint; {\em Unicast} as a solution of P2 in Section~\ref{sec:optimization}; and {\em No-Coop} for a baseline, which neither employs device-to-device connectivity nor collaboration. (We omit NUM formulation and solution of No-Coop for brevity.)

\subsubsection{Simulation Results} Next, we present the throughput results for different scenarios and policies.  We choose some of the simulated scenarios to demonstrate the key points about the performance of \MC. Additional simulation results can be found in \cite{thisAllerton}.

\begin{figure*}[t!]
\begin{center}
\subfigure[$p_{i,j}=0$]{{\includegraphics[height=6cm]{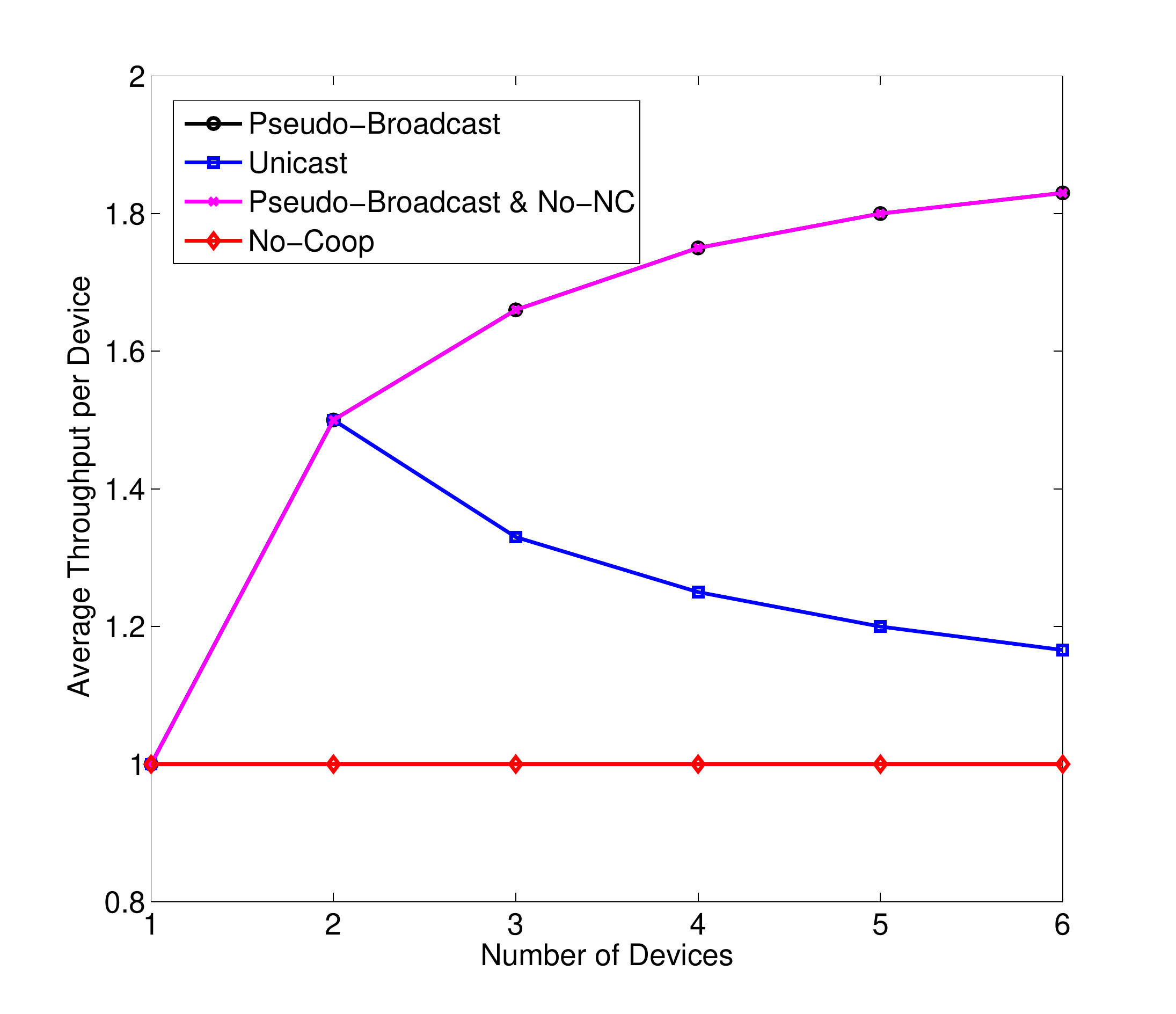}}}\hspace*{15mm}
\subfigure[$p_{i,j}=0.2$]{{\includegraphics[height=6cm]{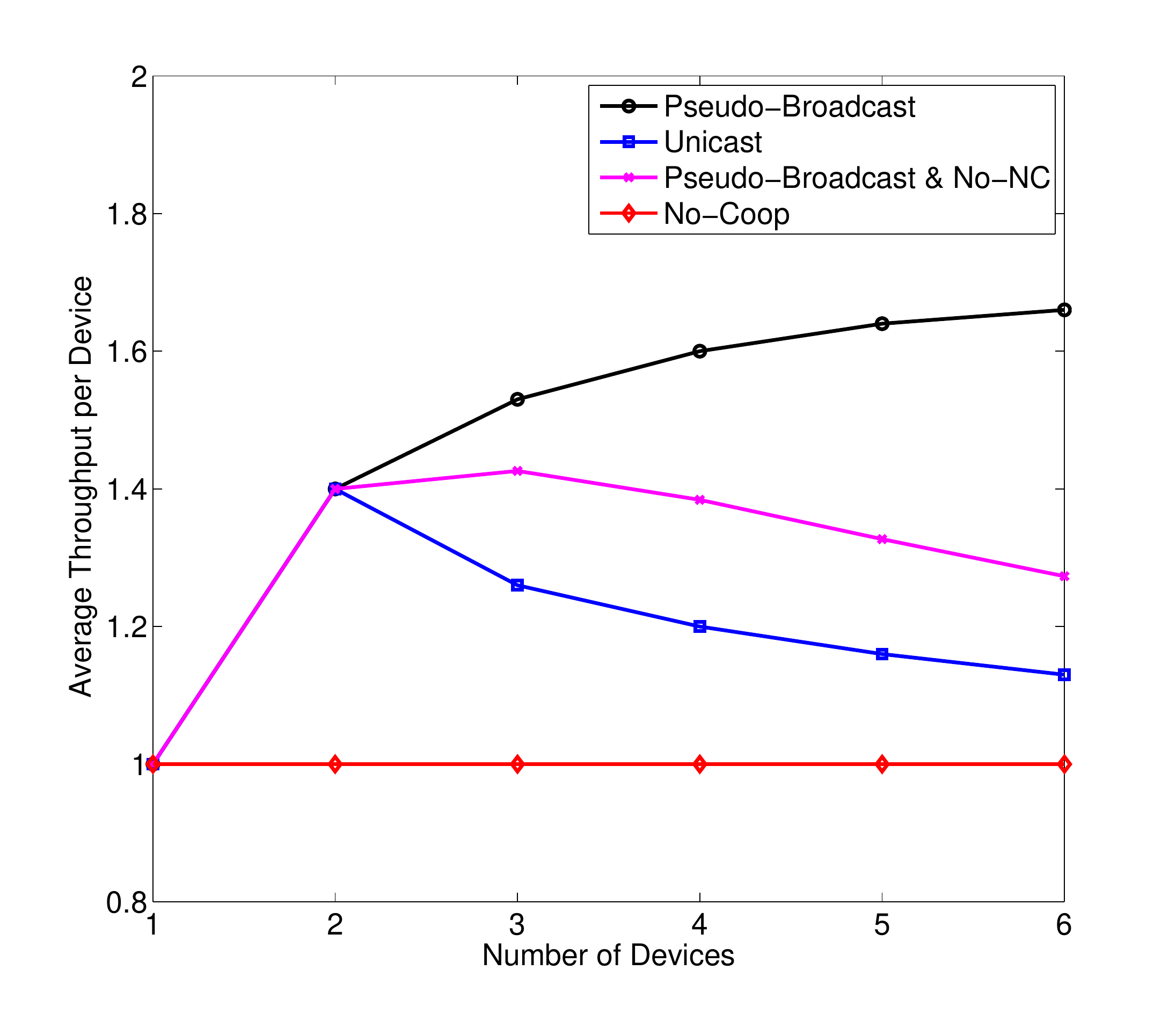}}}
\end{center}
\begin{center}
\vspace{-10pt}
\caption{\label{performance_figs_thrpt_vs_numUsers}{\bf Throughput versus number of devices.} Parameters: $C_{i}=1$, $p_{i}=0$, and $C_{i,j}=1$ for $i,j \in \{1,2,3\}$. These figures demonstrate the following points.  {\bf (1) Throughput Benefit: } \MC can increase the common throughput by a factor up to the number  of devices: in the beginning it grows linearly with the number of devices, then sublinearly, as the local network starts becoming the bottleneck. {\bf (2) Effect of Broadcast:} Local cooperation with broadcast significantly outperform local cooperation with unicast because it alleviates congestion in the local network. {\bf (3) Effect of Network Coding:} Network Coding, combined with Broadcast, helps when there is loss, by making transmissions more useful.}
\end{center}
\vspace{-30pt}
\end{figure*}

{\flushleft \bf Effect of Number of Devices.}\quad Fig.~\ref{performance_figs_thrpt_vs_numUsers} shows the average throughput  versus the number of devices for the following parameters: $C_{i}=1$, $p_{i}=0$, and $C_{i,j}=1$ for $i,j \in \{1,2,3\}$. The average throughput is calculated over all devices in the system. One can see that the throughput does not change as the number of devices increases for the No-Coop scheme. This is because there is no cooperation in the local area, so the number of devices does not affect the throughput. For {Unicast}, the throughput increases as the number of devices increases, then reduces. The reason is that cooperation in the local area helps initially. However, when the number of devices increases, the local bandwidth is no longer sufficient to support all unicast transmissions, and the unicast transmissions compete for their share of the local bandwidth. Hence, the overall throughput reduces.

The {Pseudo-Broadcast} and {Pseudo-Broadcast \& No-NC} schemes achieve the same throughput levels in Fig.~\ref{performance_figs_thrpt_vs_numUsers}(a) ($p_{i,j}=0$). Both schemes improve the throughput as the number of devices increases. The reason is that since the packets are broadcast, each transmitted packet will be beneficial to more devices when the number of devices increases. Yet, the average throughput of {Pseudo-Broadcast \& No-NC} reduces after a threshold in Fig.~\ref{performance_figs_thrpt_vs_numUsers}(b) ($p_{i,j}=0.2$). This is because when network coding is not employed, each individual packet should be successfully transmitted to all other nodes in the local area. If it is not successfully delivered, the packet is re-transmitted. This introduces inefficiency and reduces the average throughput. Meanwhile, {Pseudo-Broadcast} improves the throughput as the number of devices increases and does not introduce inefficiency thanks to network coding, which makes all packets equally beneficial.

\begin{figure*}[t!]
\begin{center}
\subfigure[$N=3$]{{\includegraphics[height=6cm]{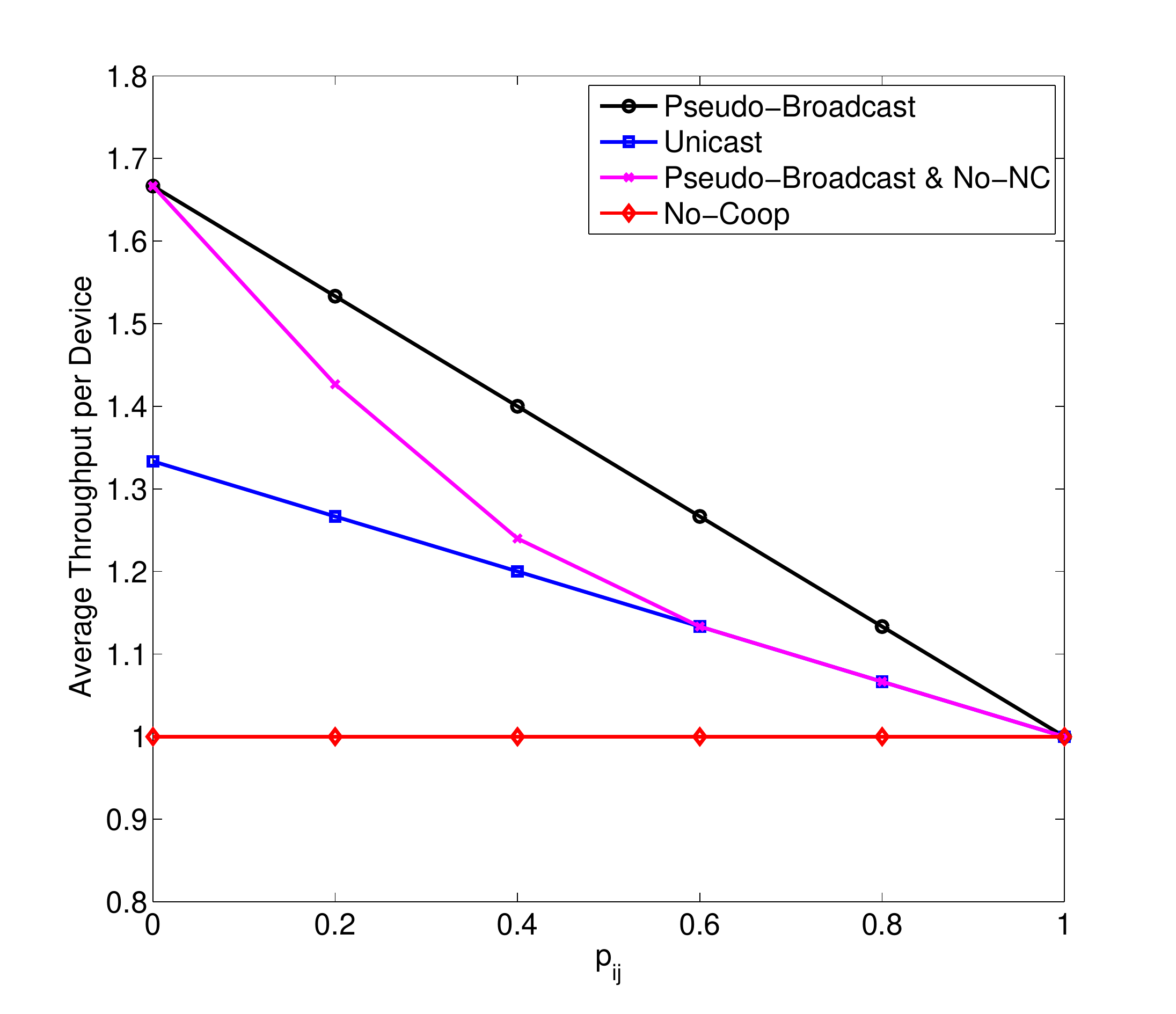}}}\hspace*{15mm}
\subfigure[$N=4$]{{\includegraphics[height=6cm]{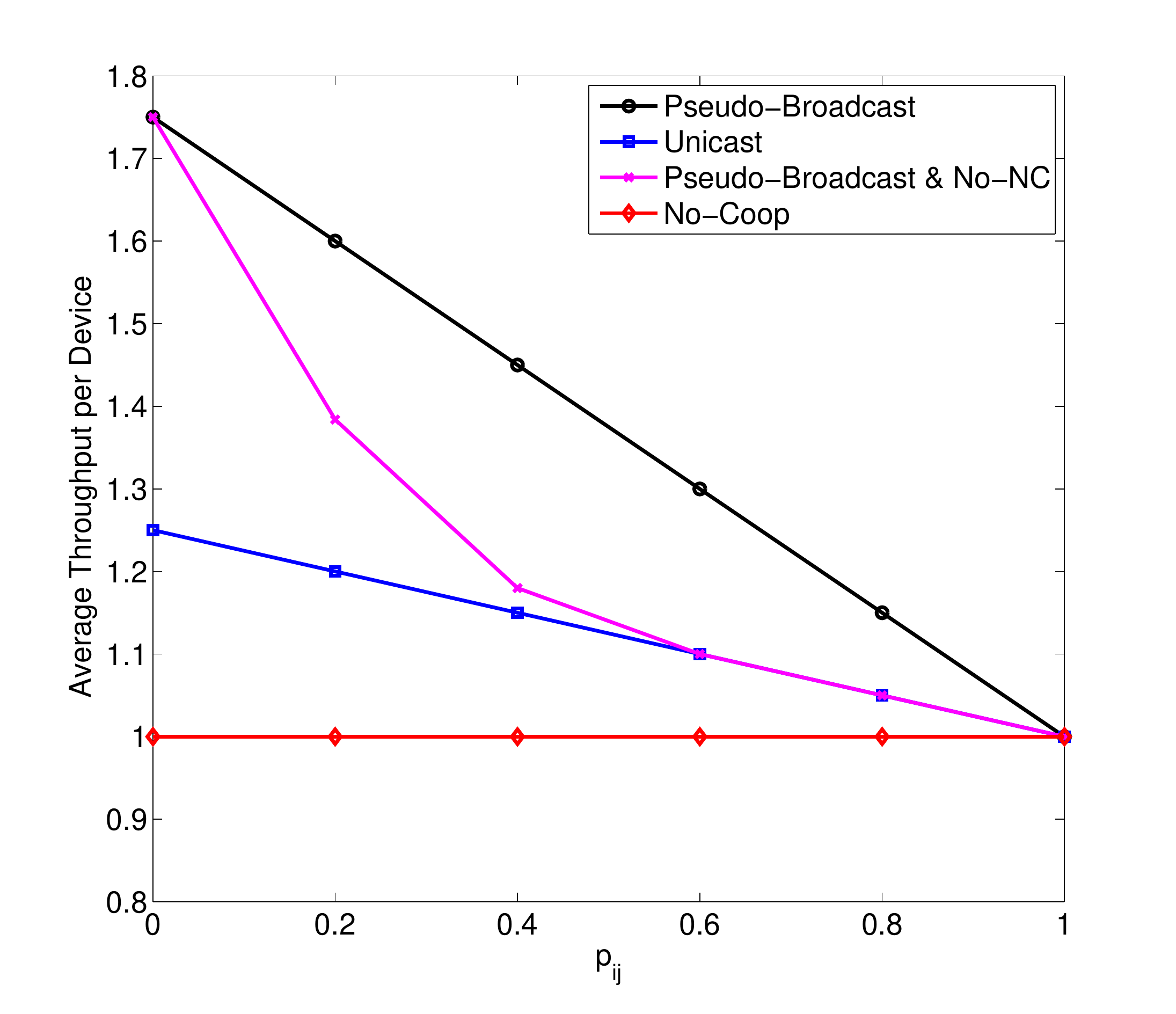}}}

\vspace{-5pt}
\caption{\label{performance_figs_thrpt_vs_pij}{\bf Throughput versus local area loss probability}. Parameters: (a) $C_i=1$, $p_i=0$, $C_{i,j}=1$, $i,j \in \{1,2,3\}$, and (b) $C_i=1$, $p_i=0$, $C_{i,j}=1$, $i,j \in \{1,2,3,4\}$. These figures show that when there is local wireless loss, Network Coding combined with Broadcast performs much better than other schemes, and the improvement is larger when there are more devices.
}
\end{center}
\vspace{-30pt}
\end{figure*}

{\flushleft \bf Effect of Wireless Loss.}\quad Fig.~\ref{performance_figs_thrpt_vs_pij} shows the average throughput versus the local area loss probability ($p_{i,j}$) for the following parameters: (a) $C_i=1$, $p_i=0$, $C_{i,j}=1$, for $i,j \in \{1,2,3\}$, and (b)  $C_i=1$, $p_i=0$, $C_{i,j}=1$, for $i,j \in \{1,2,3,4\}$. One can observe that when there is local wireless loss, both {Pseudo-Broadcast} and {Pseudo-Broadcast \& No-NC} outperform {Unicast}. {Pseudo-Broadcast} has the best performance thanks to network coding. In addition, the gap between {Pseudo-Broadcast} and {Pseudo-Broadcast \& No-NC} increases when the number of devices increases, \ie when Fig.~\ref{performance_figs_thrpt_vs_pij}(a) and Fig.~\ref{performance_figs_thrpt_vs_pij}(b) are compared. These show the benefit of using network coding and broadcast in the presence of local wireless loss.

\subsection{Experimental Evaluation of \MC}
\label{subsec:sys-eval}

In this part, we perform an experimental evaluation of the system using the Android prototype of \MC. First, we evaluate the performance of key components alone, namely \Requester and \Distributor, and compare their performance to baseline  popular alternatives.  We then evaluate the entire {\MC} system as a whole.  We show that the proposed system significantly improves the streaming experience in terms of performance (download time and video rate), without introducing significant energy cost.

We perform experiments on an  Android testbed consisting of seven smartphones: four Samsung Captivate and three Nexus S. All  smartphones have a 1 Ghz Cortex-A8 CPU and 512 MB RAM. Six of them use Android Gingerbread (2.3) and one (Nexus S) uses Android Ice Scream Sandwich (4.0) as their operating systems. In all experiments involving network coding, we use the native coding scheme (described in Section \ref{subsec:micro-ncp2}) with generation size $m=25$ and packet size $n=900$ bytes. The current implementation cannot support more than 7 (or 6) concurrent devices when an Android 4.0 (or 2.3) device acts as the access point (AP). This is due to the limitation of the soft AP currently implemented in Android.

\begin{figure*}[t!]
\begin{center}
\subfigure[The cellular link rates of three smartphones in the same geographical area.]{{\includegraphics[scale=0.28]{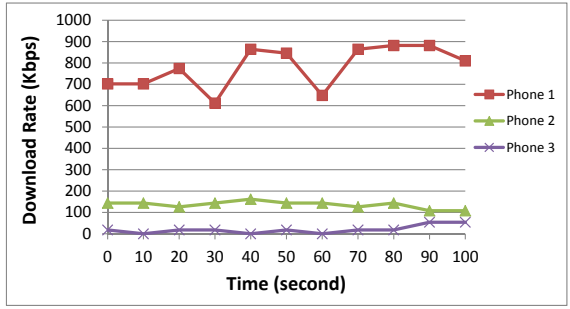}}}\hspace*{10mm}
\subfigure[The amount of local traffic introduced by four phones when using different distributors.]{{\includegraphics[scale=0.28]{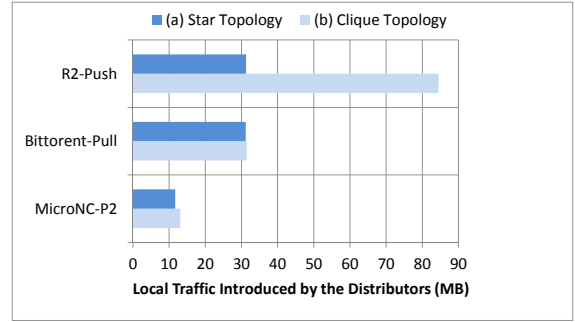}}}
\end{center}
\begin{center}
\caption{(a) {\bf Evaluation of {\Requester}:} Despite being in close proximity and being connected to the same operator, three phones experience significantly different average download rates and rate variation. This motivates the design of {\Requester} that adaptively requests data according to individual phone rates. (b) {\bf Evaluation of {\Distributor}}: A 9.93 MB file is downloaded by one of the four phones, and its segments are exchanged among the phones. This figure shows that the amount of local traffic introduced by {\Distributor} is about three times less than other schemes, thanks to Broadcast and Network Coding. \label{fig:eval-component}}
\end{center}
\vspace{-30pt}
\end{figure*}

{\flushleft \bf \Requester Evaluation.}\quad Here, we present experimental results that motivate the necessity of {\Requester} and we show its effectiveness. The setup is the following: we use three Nexus S connected to the same cellular network provider, and place them within proximity of each other (the distances among them are approximately 2 cm) in an indoor environment. The phones are placed in their positions 5 minutes before the experiment to eliminate any possible positive or negative bias due to mobility.

In our experiment, we disable \Distributor, and we measure the download rates of the smartphones over 100 seconds. The results are presented in Fig.~\ref{fig:eval-component} (a). The figure shows that despite being in close proximity and being connected to the same operator, the phones experience significantly different average download rates. Phone 3 has a very low rate because it uses EDGE. The other two phones use the same 3G network but still have significantly different download rates. Moreover, phone 1 experiences significant rate variations. This variability in time and across phones is our motivation to develop {\Requester} -- to adaptively request data according to the rates of the cellular links, instead of making static decisions, such as, splitting the requests equally among the phones.

Using these measurements, we can compare the effectiveness of \Requester to a simpler static strategy. We consider a scenario where the three phones download a $750$ kB file, and \Requester  makes a static decision: each phone requests one third of the file. In this case, phone 3 (considering the same channel realization as in Fig.~\ref{fig:eval-component} (a))  is the bottleneck for downloading the file, and the total download duration is 80 seconds. However, if \Requester makes adaptive requests, as proposed in Section~\ref{subsec:micro-download}, then phone 3 is not a bottleneck anymore, and the total download duration is less than 10 seconds. This shows the importance of the adaptive request mechanism of \Requester.

\begin{figure*}[t!]
\centering
  \subfigure[Space-Time Diagram of \Pull]{\label{fig:space-time-pull}\includegraphics[height=45mm]{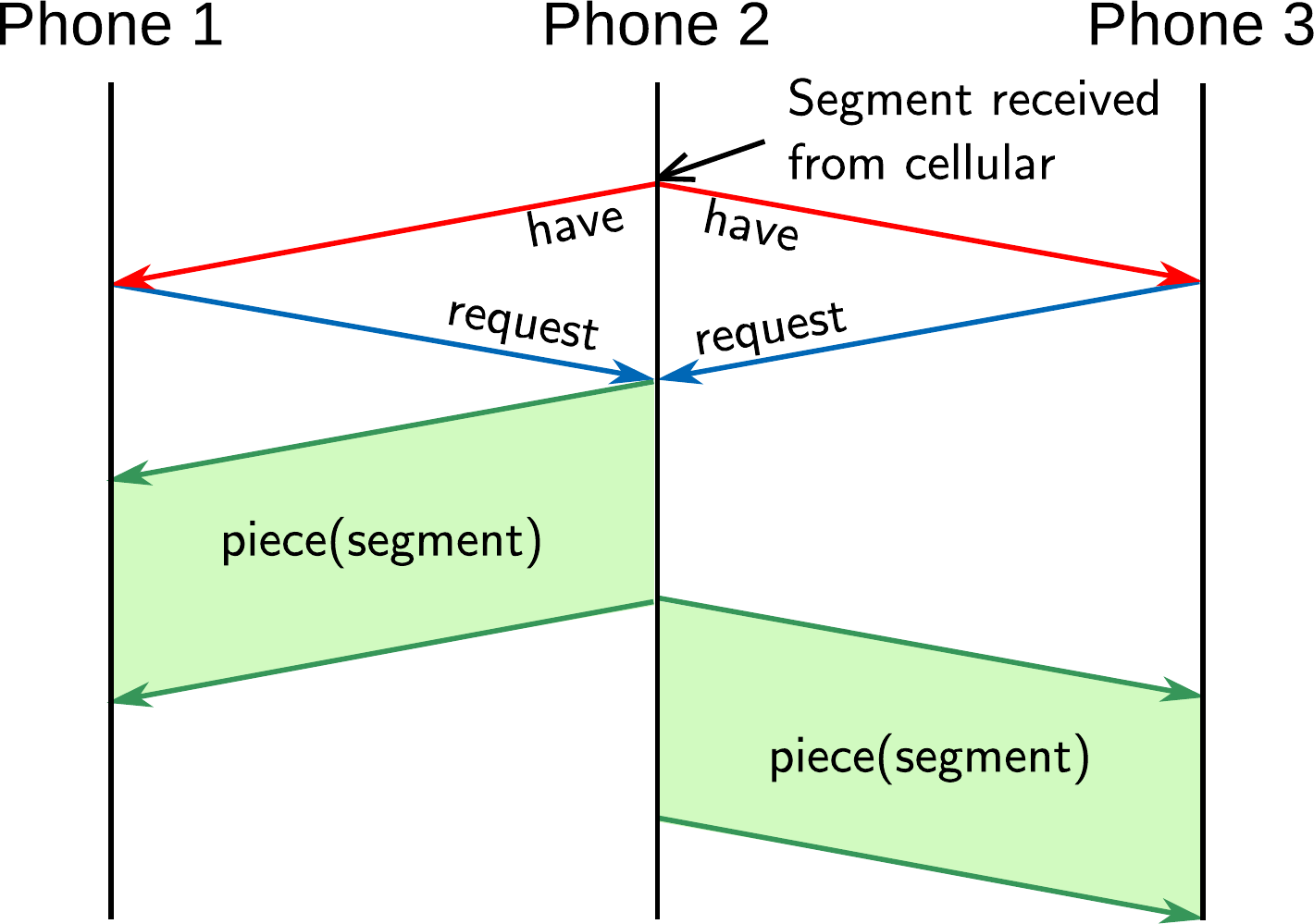}\hspace*{10mm}}
  \subfigure[Space-Time Diagram of \Push]{\label{fig:space-time-push}\includegraphics[height=45mm]{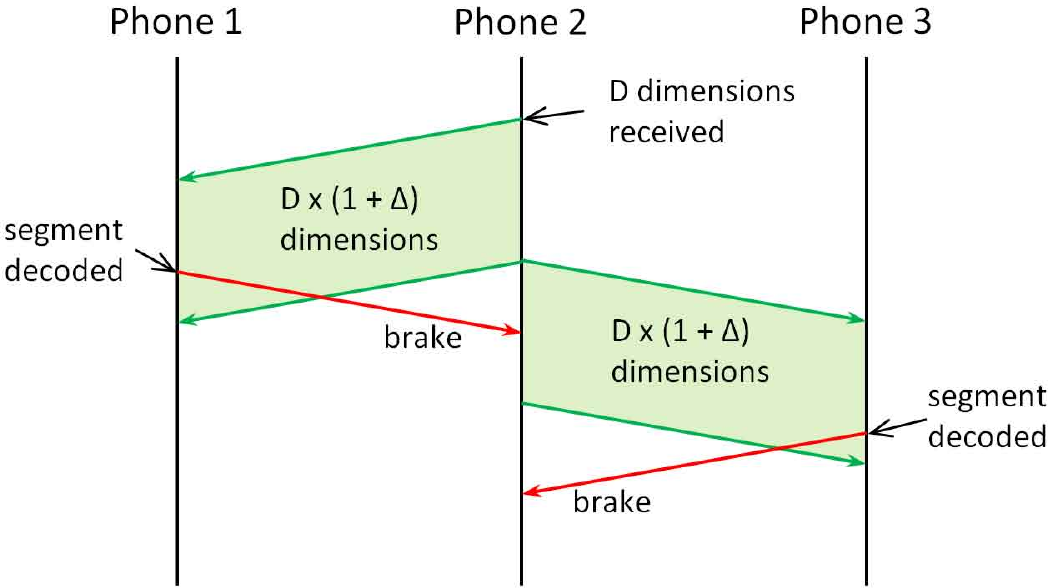}}
\caption{(a) {\bf \Pull:} When a phone has downloaded a new segment, it advertises this segment to a neighbor; the neighbor then explicitly requests the segments that it is missing; the phone then sends the requested segments. (b)  {\bf \Push:} A phone starts pushing coded packets to a neighbor as soon as it receives some packets from either the cellular network or the other neighbors. The pushing stops when a certain number of coded packets has been sent or a brake message is received.}
\label{fig:space-time}
\vspace{-20pt}
\end{figure*}

\label{evaluation_distributors}

{\flushleft \bf \Distributor Evaluation.}\quad Here, we compare the performance of \Distributor to a BitTorrent-based distributor \cite{bittorrent} and an R2-based distributor \cite{R2}. We refer to these two distributors as \Pull and \Push, respectively. Packets are exchanged locally using UDP. The performance metric of interest is the amount of local network traffic introduced by the phones when using different distributors to disseminate the same amount of information. 

We implement the \Pull scheme based on the description of the BitTorrent protocol \cite{bittorrent}. In particular, our implementation of the protocol supports three main types of messages: (i) {\em bitfield} and {\em have} messages, which are used by a phone to advertise the segments to its neighbors; (ii) {\em request} messages, which are used by a phone to request specific segments from its neighbors; and (iii) {\em piece} messages, which contain the actual data. The space-time diagram of \Pull is provided in Fig.~\ref{fig:space-time-pull}. Fundamentally, BitTorrent is a pull-based P2P protocol: when a phone has downloaded a new segment, it advertises this segment to its neighbors. The neighbors then explicitly request the segments that they are missing. To account for the wireless loss (when using UDP), we implement a recovery thread which periodically re-requests missing segments.

We implement the \Push scheme based on its description in \cite{R2}. The R2 protocol was introduced to exploit the benefit of random network coding and random push. Following \cite{R2}, our implementation of \Push supports two main types of messages: (i) {\em data} messages, which are random linear combinations of packets belonging to the same segment; and (ii) {\em brake} messages, which are used by phones to inform their neighbors that they successfully received and decoded specific segments. The purpose of brake messages is to ensure that the neighbors would stop pushing (unnecessary) linear combinations of the decoded data segments.

The space-time diagram of \Push is provided in Fig.~\ref{fig:space-time-push}. In contrast to \Pull, with \Push, the phones start pushing linear combinations of packets as soon as they receive them from either the cellular network or their neighbors. In our implementation, for a particular segment a phone is downloading, we limit the number of linear combinations that it can push to its neighbors to the rank of the matrix formed by the received packets plus a fixed amount of redundancy, $\Delta$, to account for the wireless loss rate.

Fig.~\ref{fig:eval-component} (b) shows the total amount of traffic introduced to the local network by four phones to disseminate a file when the phones are connected using star and clique topologies. In the star topology, all phones connect to the access point phone. Meanwhile, in the clique topology, all phones connect to any other phone. In both topologies, all phones overhear all the transmissions in the group. \Distributor utilizes pseudo-adhoc as described in Section~\ref{subsec:micro-broadcast}. The file size is 9.93 MB and is downloaded by a single phone using its 3G connection. The average rate of the 3G connection is measured at 550 Kbps. The phone that downloads the file is chosen at random. For \Push, we choose $\Delta = 3\%$. The local network can support 20 Mbps UDP traffic, measured using {\em iperf} \cite{iperf}. This bandwidth is much larger than what is needed to support the traffic introduced by the phones in both topologies. Since the local network bandwidth is sufficient, each of the phone receives at the rate similar to the phone which downloads the file through 3G. Each reported number is averaged over three experiments.

We first observe from Fig.~\ref{fig:eval-component} (b) that the amount of traffic introduced by both \Pull and \Push are more than three times higher than that of \Distributor. Intuitively, this is due to the fact that when using \Distributor, a packet sent by a phone may be beneficial to three phones instead of one thanks to network coding and overhearing.  Fig.~\ref{fig:eval-component} (b) also shows that in a clique topology, \Push introduces much more traffic as compared to the star topology. This is because in a clique topology, a phone may simultaneously receive linear combinations of the same segment from multiple neighbors. When this happens, it is critical that the neighbors which are sending to this phone stop pushing linear combinations in a timely manner. This could only be achieved with a timely arrival of the brake (stop) messages, which is not always possible in the clique topology, or in a setup where additional traffic is very high. The authors of R2 also observed the problem and reported it in \cite{Usee}.

In summary, the set of experimental results presented above show that by exploiting network coding and the broadcast nature of the wireless medium, \Distributor manages to introduce less amount of traffic into the local network as compared to \Pull and \Push.

\begin{figure*}[t!]
\centering
\subfigure[Average download rate as a function of number of phones in a non-congested local network.]{\includegraphics[width=5.2cm]{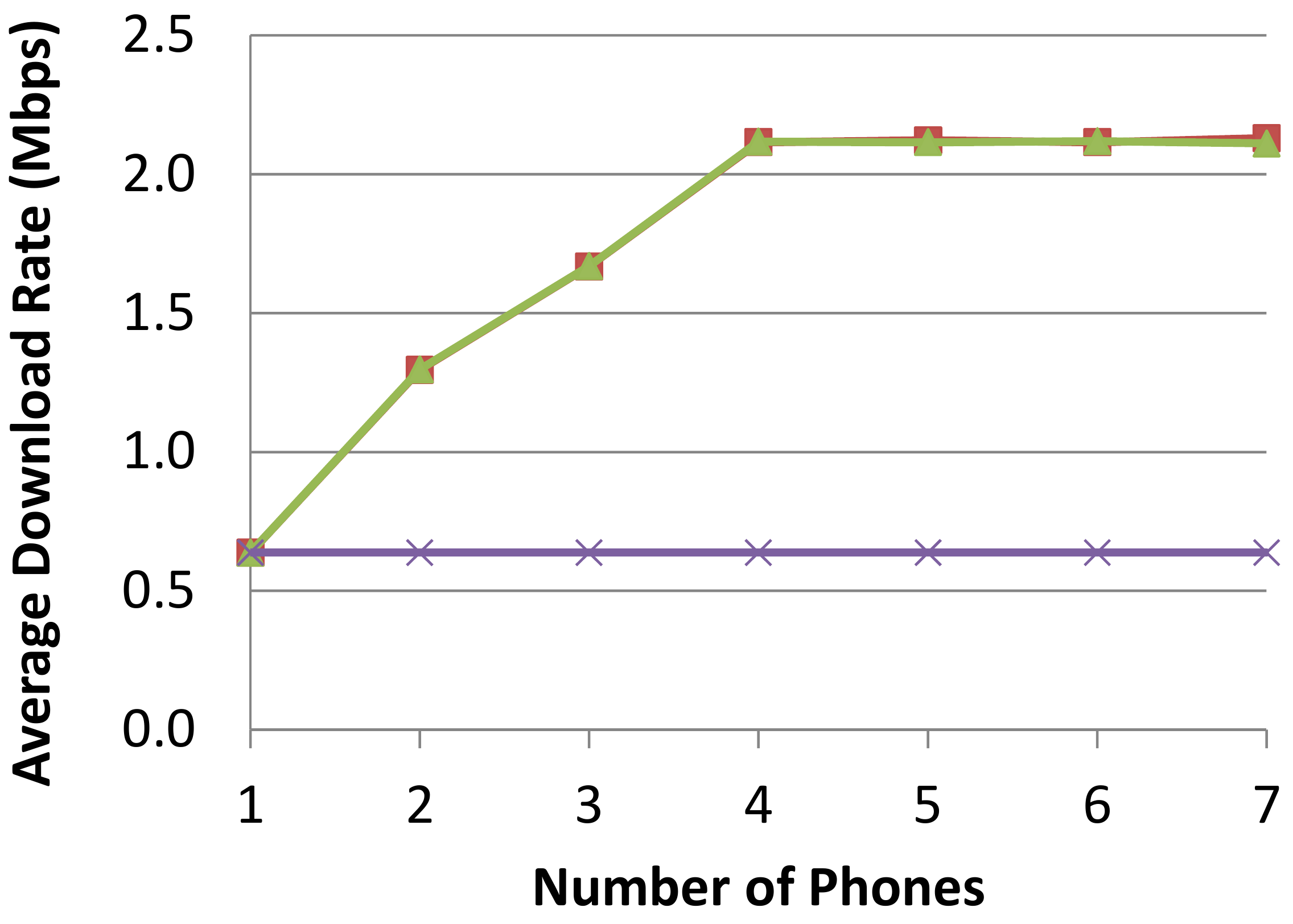}}\hspace{4mm}
\subfigure[The amount of local traffic introduced by all phones in a non-congested local network.]{\includegraphics[width=5.2cm]{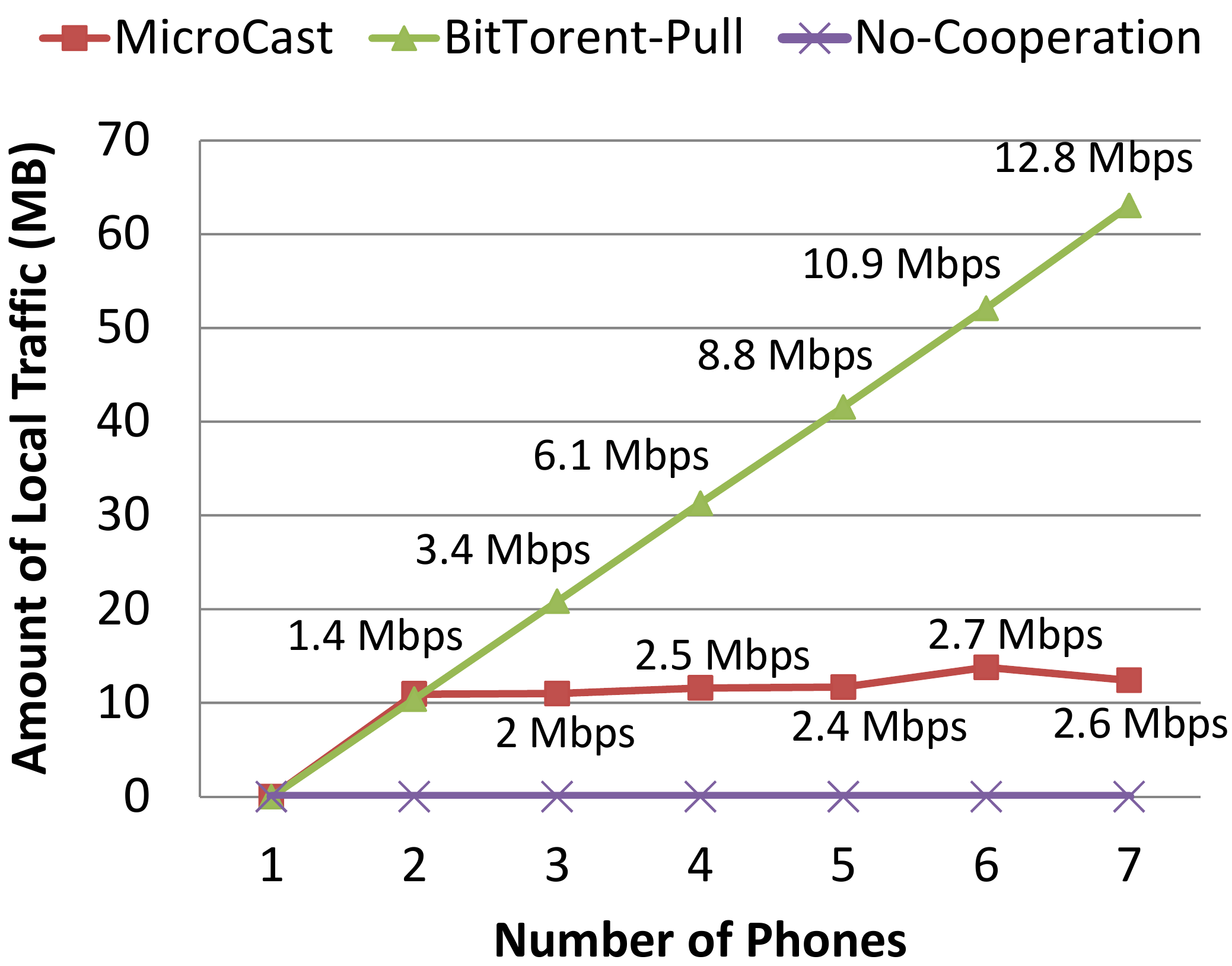}}\hspace{4mm}
\subfigure[Average download rate as a function of number of phones when the local network bandwidth is less than 4 Mbps.]{\includegraphics[width=5.2cm]{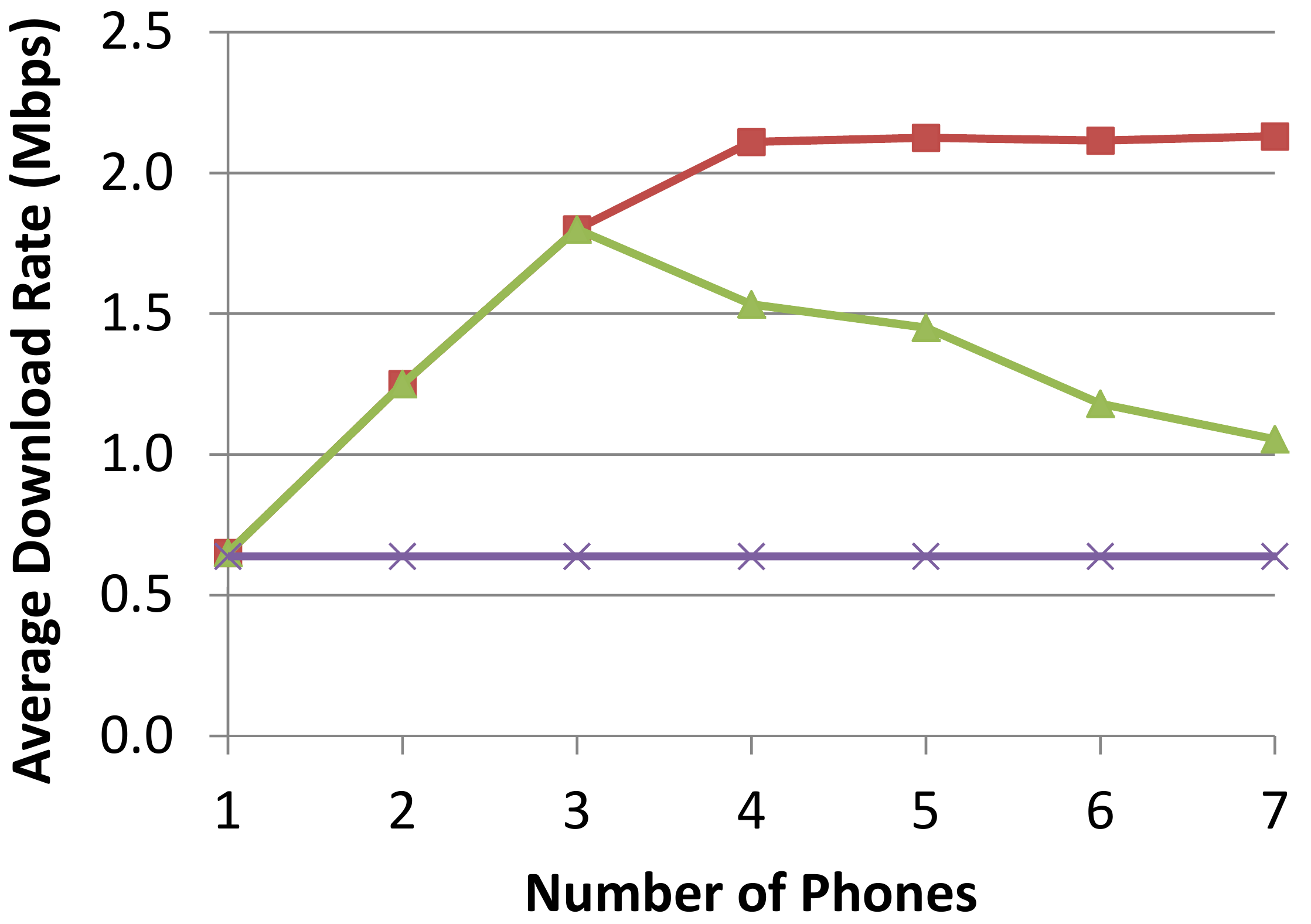}}
\begin{center}
\caption{{\bf Evaluation of \MC.} Up to 7 phones, 4 of which have 3G, cooperatively download a 9.93 MB file using different schemes. Although Fig. (a) shows similar average download rates for both \MC and \Pull,  Fig. (b) reveals that \Pull introduces a much larger amount of local traffic, which is detrimental in terms of the average download rate in congested local networks. Indeed, Fig. (c) shows that \MC still improves the average download rate up to the total capacity of the 3G links (of four phones) in a congested network, while the download rate of \Pull reduces when there are more than three phones, due to congestion. Observe that practice agrees with theory: Fig (c) is consistent with the results presented in Fig.~\ref{performance_figs_thrpt_vs_numUsers}. \label{fig:eval-MC}}
\end{center}
\vspace{-30pt}
\end{figure*}

{\flushleft \bf \MC Evaluation: Throughput.}\quad Here, we present the performance evaluation of the entire \MC system. We compare the average download rates of \MC to two other schemes: no cooperation, which we will refer to as No-Cooperation, and the combination of \Requester and \Pull, which we will simply refer to as \Pull. Note that we do not include \Push as a baseline in this section due to its inefficiency in our setup, as explained above.

In our experimental setup, we use up to seven phones. The first four phones have 3G rates varying from 480 Kbps to 670 Kbps; the rest of the phones do not have 3G connections. Packets are exchanged locally using UDP. The local network can support up to 20 Mbps UDP traffic. We use the star topology and pseudo-adhoc for \MC, and we use the clique topology for \Pull. The size of the video file is 9.93 MB. Each value reported is averaged over three experiments.

Fig.~\ref{fig:eval-MC} (a) shows the  average download rate versus the number of phones. We observe that both \MC and \Pull are able to improve the average download rate up to the total capacity of the 3G links. Note that \MC and \Pull do not provide any improvement for more than four phones because only four phones have 3G connections. Fig.~\ref{fig:eval-MC} (b) shows the amount of local traffic versus the number of phones. Although  in Fig.~\ref{fig:eval-MC} (a) we see similar average download rates for both \MC and \Pull, Fig.~\ref{fig:eval-MC} (b) shows that \Pull introduces a much larger amount of local traffic, which increases linearly in the number of phones, when compared to \MC. This behavior of \Pull is detrimental in terms of the average download rate in congested networks. An important observation is that, as the number of phones increases, \MC's local traffic does not increase. This indicates that, even when there are many phones, overheard packets are lost very rarely.

We then update our experimental setup to evaluate the performance of \MC and \Pull in a congested network. In our new setup, the congested network is generated by introducing 16 Mbps background UDP traffic on the same 802.11 channel. (Note that there is also interference from other sources in the environment which contributes to the background traffic.) Since the local network can support up to 20 Mbps traffic, the leftover traffic is less than 4 Mbps. Fig.~\ref{fig:eval-MC} (c) presents the average download rate versus the number of phones in this setup. We see that the average download rate  of \Pull  reduces when we have more than three phones. This is because \Pull introduces a large amount of local traffic (as illustrated in Fig.~\ref{fig:eval-MC} (b)), which leads to congestion.  Note that the addition of the fifth, sixth, and seventh phones also increases the local traffic in \Pull even though they do not have 3G connection. This is because they still need to receive the file in the local area, which contributes to the local area traffic.

In contrast, Fig.~\ref{fig:eval-MC} (c) shows that \MC still improves the average download rate up to the total capacity of the 3G links (of four phones) in a congested network. This is because it introduces only a small amount of local traffic (\eg even for seven phones, \MC only introduces 2.6 Mbps traffic to the local network as in Fig.~\ref{fig:eval-MC} (b)). It can be observed from Fig.~\ref{fig:eval-MC} (c) that the average download rate of \MC is more than three times higher than that of No-Cooperation. Also, the improvement of \MC over \Pull in terms of average download rate is as high as 75\% (we observed even more improvement for different setups, \eg with 18 Mbps background traffic), which is significant.

The result demonstrated in Fig.~\ref{fig:eval-MC} (c)  is consistent with the simulation results presented in Section \ref{subsec:num-eval} (Fig.~\ref{performance_figs_thrpt_vs_numUsers}). In practice, for the \Pull (and the unicast policy), the point at which adding an additional device hurts the average download rate depends on a number of factors, mainly how congested the local network is currently, the cellular rates of the existing devices, and the video rate. As shown in Fig.~\ref{performance_figs_thrpt_vs_numUsers}, the adding of the third device to the group immediately reduces the average download rate; meanwhile, in Fig.~\ref{fig:eval-MC} (c), this does not happen until the addition of the fourth device.

\begin{figure*}[t!]
\begin{center}
\subfigure[Battery drain when downloading the same file of size 95.4 MB using different schemes.]{{\includegraphics[scale=0.28]{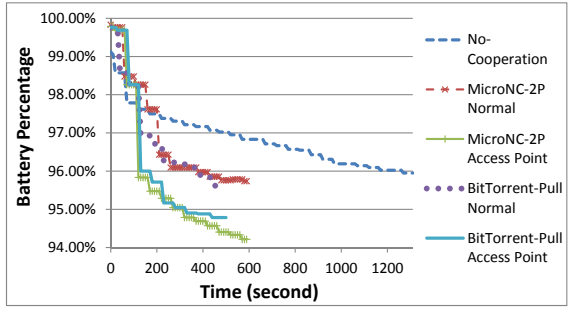}}}\hspace*{10mm}
\subfigure[Coding and decoding throughput as a function of the generation size. Each packet is 900-byte long.]{{\includegraphics[scale=0.28]{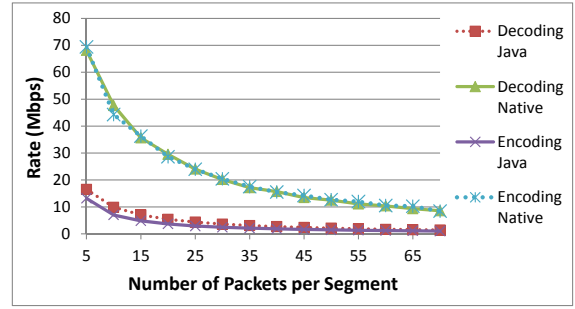}}}
\end{center}
\begin{center}
\caption{Beyond video throughput. (a) {\bf Evaluation of energy consumption:}  No-Cooperation consumes less battery per unit time but takes longer to download the file compared to the other. Battery levels of all schemes are similar when the file transmission is completed. This demonstrates that employing cooperation, in the long term does not bring any significant battery cost. (b) {\bf Evaluation of network coding implementations:} Observe that the coding rate is much higher when using the native implementation. The figure shows that it is possible to support high rate network coding on commodity phones -- more than sufficient to support current standard video streaming rates.
\label{figs:eval-energy-coding}}
\end{center}
\vspace{-30pt}
\end{figure*}

{\flushleft \bf \MC Evaluation: Energy Consumption.}\quad We compare the energy consumption of \MC with the baselines: \Pull and No-Cooperation, when downloading a video file. We consider a star topology similar to the one above: three phones connect to the fourth one that acts as the access point (AP). All phones have 3G with rates varying from 450 Kbps to 700 Kbps, and the size of the video file is 95.4 MB. We use the {\em BatteryManager} class of the  Android SDK for the power consumption measurements. Before the experiment, all four phones are fully charged. During the experiment, the battery states are recorded  every 10 seconds. The experiments are repeated three times, and their average is reported.

Fig.~\ref{figs:eval-energy-coding} (a) presents the battery state (100\% corresponds to the fully charged battery) versus time. Note that ``\Distributor Access Point'' and ``\Pull Access Point'' show the battery consumption levels of the phones selected to act as the APs in \MC and \Pull, respectively. Meanwhile, ``\Distributor Normal'' and ``\Pull Normal'' show the battery consumption levels of phones which are not the APs. It can be observed from Fig.~\ref{figs:eval-energy-coding} (a) that the No-Cooperation scheme has less battery consumption compared to \MC  and \Pull at any given time within 0 and 600 second. However, the time required to download the video file when using No-Cooperation is very high (more than two times as compared to \MC and \Pull). 

Observing the battery levels of all schemes when the file transmission is completed, one can see that the battery consumption levels of No-Cooperation, \MC, and \Pull are similar. This demonstrates that employing cooperation, in the long term (to download a video file), does not bring any significant battery cost. Finally, the phones which act as APs consume more battery than the other phones. This is expected because the AP phone has to perform additional tasks (scheduling packets, maintaining connection, etc.). Nevertheless, even in the worst case, for the AP phone, \MC consumed approximately 6\% of the battery to download the whole file. Considering that the phones are downloading a large file (95.4 MB), this battery consumption level is reasonable. These considerations show that the rate benefit of \MC comes at no significant battery cost.

{\flushleft \bf Network Coding Evaluation.} We evaluate the performance of the two implementations of network coding we developed: native (written in C) and Java-based coding. The two implementations are described in Section~\ref{subsec:micro-ncp2}. Fig.~\ref{figs:eval-energy-coding} (b) shows the decoding and encoding data rates as a function of the generation size. The slowest encoding rate for the Java implementation is 1 Mbps, while for the native implementation, it is 8 Mbps. Fig.~\ref{figs:eval-energy-coding} (b) also shows that coding rate is much higher when using the native implementation. In conclusion, we find that it is possible to support high rate network coding on commodity phones, more than sufficient to support current standard video streaming rates (a standard YouTube 480p video rate is 2.5 Mbps  \cite{youtube-rate}).

\section{\label{sec:conclusion} Conclusion}

In this work, we proposed a cooperative system for video streaming to  a group of mobile device users, who are within proximity of each other, and are all interested in viewing the same video at roughly the same time. The proposed system is grounded on a NUM formulation and its distributed solution. The system cooperatively uses the network resources on all devices of the group, such as, cellular and WiFi links,  to improve the streaming experience. We evaluate the proposed system through both simulation and experimentation by implementing a full working prototype on the Android platform. Evaluation results  demonstrate significant performance benefits in terms of per-user video download rate without significant battery penalty. Additional materials and video demonstration can be found at \cite{projectWiki}.

\balance
\bibliographystyle{IEEEtran}
\bibliography{main.bbl}

\end{document}